\documentclass[pdflatex,sn-apa]{sn-jnl}

\pdfoutput=1

\usepackage{xspace}
\usepackage{amssymb}
\usepackage{amsmath}
\usepackage{graphicx}
\usepackage{url}
\usepackage{hyperref}
\usepackage{natbib}

\newcommand\lhood{{\cal L}}
\newcommand\cl{{\textnormal{CL}}\xspace}

\newcommand\mulo{\mu_\textnormal{L}} %lower end of two-sided conf. intervals
\newcommand\muhi{\mu_\textnormal{U}} %high end of two-sided conf. intervals
\newcommand\muUL{\mu_\textnormal{UL}} %upper limit
\newcommand\muLL{\mu_\textnormal{LL}} %lower limit

\newcommand\pcl{{\ensuremath {p}}{\mbox{-CL}}\xspace}

\newcommand{\boldX}{\ensuremath{\textbf{\textit{X}}}\xspace}
\newcommand{\boldx}{\ensuremath{\textbf{\textit{x}}}\xspace}

\newcommand{\xzero}{\ensuremath{x_0}\xspace}
\newcommand{\bxzero}{\ensuremath{\textbf{\textit{x}$_0$}}\xspace}

\newcommand{\dofx}{\ensuremath{\mathrm{d}(\mathbf{x})}\xspace}

\newcommand{\rxzero}{\ensuremath{\mathrm{r}(\mathbf{x}_0)}\xspace}

\newcommand{\rofx}{\ensuremath{\mathrm{r}(\mathbf{x})}\xspace}
\newcommand{\rofxNOBOLD}{\ensuremath{\mathrm{r}(\mathrm{x})}\xspace}

\newcommand{\PR}{\ensuremath{\mathrm{\Pr}}\xspace}

\newcommand\cls{CL$_\mathrm{s}$}

\begin{document}

\begin{center}
    {\LARGE\bfseries Examples of statistical practice in elementary particle physics for comparison with Mayo's severity concept\par}
    
    \vskip 1.5em
    
    {\large
    Robert D. Cousins\\
    Dept.\ of Physics and Astronomy\\
    University of California, Los Angeles\\
    Los Angeles, California 90095\\
    \texttt{cousins@physics.ucla.edu}
    \par}
    
    \vskip 1em
    
    {\normalsize October 7, 2026\par}
\end{center}

\begin{abstract}
\\ The field of elementary particle physics (also known as high energy physics) has strong traditions in frequentist inference, heavily influenced by Fisher and by Neyman and Pearson. While I am not aware of direct use and citation of Deborah Mayo's severity concept in the field, one might expect that ideas similar to severity were independently developed. Here, I describe typical practice that addresses the same issues that severity was developed to address, with some tentative comparisons with severity. I invite experts on severity to explain how they would apply severity in these cases and to discuss any differences.
\end{abstract}

\newpage

\section{Introduction}
In elementary particle physics (also known as high energy physics, HEP), the frequentist approach (called ``classical" for some decades) is rooted in long traditions, including influential books such as \citet{eadie1971}, in turn based heavily on the treatises of 
Kendall and Stuart and successors \citep*{kendall1999}.
The widely used software tool Minuit~\citep{JamesMinuit1975,james1980} facilitated the use of the likelihood function for point and interval estimation, the latter using both the curvature of the log-likelihood function and differences in log-likelihood (interpreted using the theorem of \citet{wilks38}). In Minuit, nuisance parameters are eliminated by the method that we eventually learned was called ``profile likelihood" in the statistics literature~\citep{rolkeclk}.

The processes that we study are governed by quantum mechanics, which is viewed by almost everyone in the field as providing the means for obtaining samples that, at their core, are perfectly independent and identically distributed, even if viewed through the imperfect and not identical capabilities of detectors and transducers.  The modestly named `Standard Model (SM) is a mature theoretical framework that successfully provides detailed descriptions of a wide range of particles and their interactions, while also leaving many open questions that are the subject of current research.

Frequentist hypothesis tests in the tradition of Neyman and Pearson (N-P) are routinely performed, accompanied by reporting of post-data $p$-values in the tradition of Fisher, and typically mapped to $Z$ values (i.e., number of equivalent Gaussian standard deviations $\sigma$). I do not think that there was ever a time when confidence intervals were not also routinely reported (corresponding to what are known in psychology as ``effect size in original measurement units"). Indeed, interval estimation (often called ``error analysis") has been a feature of first-year university laboratory exercises since before my time.

Several categories of hypothesis tests are commonly performed in HEP, including:
\begin{description}
\item{(a)}
A given functional form (``model'') vs.\ a second functional form;
\item{(b)}
A specified value of a parameter (say 0) vs.\ another specified value of that parameter (say 1), with essentially the same techniques as used for binary classification;
\item{(c)}
A single value of a parameter (say 0 or 1) nested within a continuous set of alternative values;
\item{(d)}
A given functional form against all other
(unspecified) functional forms (goodness of fit).
\end{description}

I assume that the reader has general knowledge of traditional methods for performing such tests in the frequentist context, including concepts associated with N-P and/or Fisher and their followers, such as Type I error probability $\alpha$; Type II error probability $\beta$; the power 1$-$$\beta$; as well as likelihood functions and ratios (as used by both Fisher and N-P). 

My discussion here focuses on cases (b) and (c). As some of us first learned over 25 years ago from \citet[Section 20.9]{kendall1999}, the hypothesis testing case (c) is mathematically equivalent to interval estimation. That is, as discussed more thoroughly in Section~\ref{background}, a confidence interval for $\mu$ at confidence level 1$-$$\alpha$ consists of all values of $\mu$ that are {\em not} rejected by a hypothesis test of Type I error probability $\alpha$.  

This duality of ``inversion" of a hypothesis test to get confidence intervals, and vice versa, is fundamental. I refer to the two sides of the duality as the {\em interval picture} and the {\em testing picture}.  While the duality is always there, in some contexts in HEP we tend to think almost exclusively in terms of the interval picture, and in some other contexts the testing picture.

It is understood in HEP that the accept/reject nomenclature of N-P does not constitute a complete formal theory of {\em decision making}, i.e.,  what action should I take (tell no one, issue a press release, propose a new experiment,\dots) based on the observed data.  In fact, decision making is rarely, if ever, performed formally in HEP. I would say that important high-level decisions incorporate all the inputs of a complete decision theory (the model(s), the data, prior belief, loss or utility, etc.), while not respecting all the elements or algorithms of a formal Wald or Bayesian theory. 

Our traditions for communicating numerical experimental results do not involve incorporating personal prior probabilities or utility functions. Of course, the discussion sections of papers (and abstracts) often do comment on the scientific importance of a result, whether it was expected or surprising, and how it constrains or supports speculative ideas. While physicists in HEP have long been aware of the usefulness of elements of N-P theory, e.g., specifying alternative hypotheses, giving consideration to pre-data Type I and Type II errors, they are also mindful of pitfalls of excessive rigidity such as reporting only ``accept" or ``reject"). 

In category (b) above, it is helpful to distinguish between two situations.  The first is binary classification in computer algorithms, where (for example) the goal is to distinguish one type of particle (say a pion) from another type of particle (say a muon), in many different IID beam-target reactions.  
This sort of ``industrial" application falls into the behavioristic rationale and has long been characterized by plots equivalent to modern ROC curves. Once the operating point is chosen (with well-understood Type I and Type II errors), there is little concern with individual $p$-values.

The second situation in category (b) is at the end of a high-level analysis, when one is combining all the data to test a physics hypothesis vs.\ an alternative.  An example is discussed in detail in Section~\ref{higgsCP}.  In such situations, the inferential rationale is probably a better description, with the focus on $p$-values and confidence intervals (and implicit confidence distributions), without rigid adherence to nominal conventions for previously defined thresholds. However, I note that ~\citet{neyman1937} emphasized that the frequency properties of confidence intervals (coverage, dual to errors on testing) are applicable to the ensemble of {\em different} measurements with different models, not just to the (imagined) ensemble of measurements using the same model. This may, to some extent, be the rationale in the mind of those who might think that ``5$\sigma$ is a discovery", an attitude that I find too simplistic~\citep{cousinsJL} and not really the deep thought process of professionals (regardless of what is in popular accounts).

In any case, in my experience, most physicists in HEP do not venture too deeply, if at all, into philosophical issues such as the behavioristic rationale for testing versus the inferential rationale. Essentially everyone is well-schooled in running Monte Carlo (MC) simulations and viewing multitudes of histograms that use probabilities arising from quantum mechanics, so that the frequency interpretation of coverage, tail probabilities such as Type I/II errors, and their statistical fluctuations, is routine.

Following the publication of Deborah Mayo's monograph, {\em Statistical Inference as Severe Testing: How to Get Beyond the Statistics Wars}~\citep[referred to here as SIST]{SIST}, I made a serious attempt to understand its comprehensive exposition of arguments (old and new) in favor of frequentist statistics. The concept of severity that supplements the more widely known frequentist methods was new to me (and remains largely unexplored in HEP).  As the present paper is intended to be part of a collection that includes an introduction to severity, in the interest of space, I just quote the specification from p.\ 92 of SIST:

\medskip\noindent
``{\em Severity Requirement:} for data to warrant a hypothesis $H$ requires not just that
\begin{description}
    \item{(S-1)} $H$ agrees with the data ($H$ passes the test), but also
    \item{(S-2)} with high probability, $H$ would not have passed the test so well, were $H$ false."
\end{description}
For more discussion, in addition to the papers by Mayo and others cited here, I also recommend the short, readable introduction in Section 9.4 of the book by Kent \citet{staley}.

Given the highly developed level of frequentist data analysis in HEP (assisted since the year 2000 by statisticians participating in the \cite{phystat} series of workshops organized by Louis Lyons and others), my main thought in approaching the severity concept was, ``If it is really useful, we are probably already doing something equivalent.'' 

I initially found that to be the case, with the key connection being confidence intervals that have an adjustable post-data critical confidence level (Section~\ref{missing}), in the same manner that the $p$-value is a post-data critical value of the significance level $\alpha$.  However, as pointed out by Samuel~\citet{fletcher}, my implementation of the connection between severity and the post-data critical confidence level was based on an incorrect understanding of severity, even if the numerical results agreed in some cases. In the present paper (at least in this version), I mostly refrain from performing calculations of severity, giving priority to (already long) explanations of HEP practice.

After recently reviewing some of the literature advocating severity, I believe that methods in HEP have been developed to protect against essentially the same issues in N-P testing that severity was designed to protect against. These were enumerated by \citet*{mayospanos2006} as:
\begin{description}
\item (i) coarseness (binary output of accept or reject);
\item (ii) open to fallacies of rejection (statistical significance versus substantive significance), and 
\item(iii) focus on pre-data, behavioristic goals that may conflict with appropriate post-data inference. 
\end{description}

More recently, \citet{spanos2014} describes:
\begin{description}
\item (i) ``the {\em fallacy of rejection}: (mis)interpreting reject $H_0$ (evidence {\em against $H_0$}) as evidence {\em for} a particular $H_1$; this can arise when a test has very high power, e.g., large $n$. A number of attempts have been made to alleviate the large $n$ problem, including rules of thumb for decreasing $\alpha$ as $n$ increases; see Lehmann (1986). Due to the trade-off between the Type I and Type II error probabilities, however, any attempt to ameliorate the problem renders the inference susceptible to the reverse fallacy,"
\item (ii) ``the {\em fallacy of acceptance}:
(mis)interpreting accept $H_0$ (no evidence {\em against} $H_0$) as evidence {\em for} $H_0$; this can easily arise when a test has very low power; e.g., $\alpha$ is tiny or $n$ is too small. 
\end{description}
Spanos further states that, ``These fallacies are routinely committed by  practitioners in many applied fields. After numerous unsuccessful attempts, Mayo (1996) provided reasoned answers to these fallacies in the form of a post-data severity assessment."

The present paper is dedicated to explaining how we handle these situations in HEP. (Certainly, these fallacies are not routinely committed.) I am still working to understand severity, particularly when the null hypothesis is rejected and in cases other than the one-sided test illustrated in SIST. Thus, I would welcome further clarification regarding the comparisons with severity.

As with SIST, I consider separately the two cases of ``null hypothesis accepted" and ``null hypothesis rejected". For the former, the usual practice in HEP of quoting upper limits (at least implicitly at more than one confidence level) seems to map conceptually onto severity concepts. For the latter, comparison can be obscured by the fact that we typically perform two-tailed tests of alternative point hypotheses if the initial one-tailed test rejects the null hypothesis. In general, our measures do not seem to be numerically identical to severity in all cases, but I think that they address the issues that severity was designed to address.

Section 2 introduces measurement models, the HEP prototype test, and the duality between testing and confidence intervals. Section 3 discusses post-data quantities ($p$-values, $p$-CL, and their potential relation to severity). Sections 4 and 5 examine concrete HEP analyses. The final section contains concluding remarks.

\section{HEP models and the ``HEP prototype test"}
\label{background}

\subsection{Measurement models in HEP}

In many HEP experiments, everyday particles such as electrons or protons (and their anti-particles) are accelerated in beams to high speeds and collided with stationary targets or with other beams. In the collisions, some of the energy of motion (kinetic energy) of the particles is converted to the mass of heavier particles, always obeying the equations (known as ``kinematics") of Einstein's theory of special relativity. Most of the newly formed particles decay quickly to lighter particles including electrons, photons, and other more exotic particles (pions, muons, etc.), collectively called ``objects". Highly specialized custom-designed equipment and electronic circuits, referred to as ``detectors", record the positions and ionization deposits from the objects.  

Software programs reconstruct patterns that are identified as the various objects, along with their positions, energies, and momenta. The kinematic equations of special relativity are essentially propagated backwards in time to determine the masses of the heavy particles originally produced. An enormous effort goes into calibrating the responses of all the detectors, which requires understanding the thousands of (often nonlinear) functions that convert (say) ionization signals digitized and recorded by the detectors into point and interval estimates of kinematical properties. After correcting for all known systematic effects in these object measurements, the residual uncertainties in the object properties are considered to be ``systematic uncertainties" in the measurement models, and are represented by expressions (such as normal or lognormal distributions) containing ``nuisance parameters". Other sources of systematic uncertainty are considered as well.

The frequentist sampling properties of essentially everything we do are studied using traditional MC simulation methods (less often Markov Chain MC).  We perform simulations to obtain the distributions of each test statistic under the null and alternative hypotheses, and to evaluate frequentist coverage of interval estimates (whether obtained by frequentist or Bayesian-inspired recipes).  Asymptotic formulas such as Wilks's Theorem and its generalizations are also routinely used, though users generally check by MC simulation that the asymptotic approximations are valid, if in a novel situation.

Over the course of an experimental program (years or decades), the sample size (number of beam-target collisions) can vary by many factors of ten as both the beam intensity and the observation time are increased.  Our samples are generally in the extreme Poisson regime of binomial distributions, so the sample size determines what we call the ``statistical uncertainty", which is the familiar uncertainty in a measurement of a Poisson mean (or binomial parameter) that decreases as the inverse of the square root of the sample size.  

The combination of the systematic and statistical uncertainties is often approximately that obtained by addition in quadrature.  Great effort goes into trying to keep the overall systematic uncertainty smaller than the statistical uncertainty.  This is a never-ending program as sample size increases, and not always achieved. In addition, a lot of thought goes into the correlations among the nuisance parameters.

Much effort goes into understanding how best to deal with nuisance parameters, which are present in all our statistical models, at a minimum to represent residual uncertainties in the calibrations of responses of our instruments. At CERN's Large Hadron Collider (LHC), profile likelihood functions are a common tool for dealing with nuisance parameters, though Bayesian-style marginalization (while treating the parameter of interest by frequentist methods) is also performed. For us, the bottom line is which of these approximate methods gives the best frequentist performance for parameter(s) of interest, and under which circumstances \citep{cousinswasserman}.  In recent years, the use of MC simulations to compute distributions of profile likelihood test statistics (the parametric bootstrap) has become common.  In this paper, I suppress reference to nuisance parameters, but one can assume that they are present (and a nuisance).

Furthermore, we appreciate well the point emphasized early and often in SIST (pp.\ 20, 94, 122, and others listed in the index under ``auditing'') that the experimental procedure and the statistical model must be scrutinized, with systematic effects in particular identified and quantified.  As I wrote \citep{cousinsJL}, ``It is certainly worth trying to understand if some physical parameter in the alternative core physics model is zero (corresponding to the SM), even if it is necessary to do so through the smoke of imperfect detector descriptions with many uninteresting and imperfectly known nuisance parameters.  Indeed, much of what distinguishes the capabilities of experimenters is how well they can do precisely that by determining the detector response through careful calibration and cross-checks.'' Looking back through history at (in)famous wrong results in HEP, most of them were caused by failure to specify correctly the measurement model (both the measurement apparatus and the procedure). However, in this paper, the focus is on idealized situations where the model is correctly known (including expressions for uncertainties in nuisance parameters).

\subsection{The HEP prototype test}
\label{proto}
At the discovery frontiers of the field, where one is searching for new elementary particles and/or new forces of nature, the prototype test (idealized for this discussion) contains a parameter $\mu$ that characterizes the size of a hypothetical new phenomenon.  This can be the strength of a new force, the production rate of a new particle, or decay rates of known particles to final states that are forbidden for the known forces. In such cases, $\mu=0$ means the absence of the new phenomenon, while $\mu>0$ means ``new physics''.  There is often no firm prediction for a specific value of $\mu>0$.

Sometimes, we add the complication that $\mu$ is a vector with two (or more) components of fundamental physics interest; one component could be the strength just described, while the other component could be a particle's mass, or some other parameter in the speculative theory, such as a ``mixing angle'' that represents misalignment of sets of basis vectors in some Hilbert space of quantum field theory.  (This is common in neutrino physics, for example.) However, unless otherwise stated, in this paper $\mu$ is a single (scalar) parameter.

I let \boldx (boldface following SIST) denote observable(s). More generally, \boldx is any useful function of the observable(s), i.e., a ``statistic'' or ``test statistic''. Following usage in HEP as well as in SIST (p.\ 34, etc.), we let $\bxzero$ denote a particular observed value (rather than using the statistician's notation \boldX and observed \boldx).  As with $\mu$, I sometimes include the case where \boldx is a vector with two or more qualitatively different components. E.g., one component could be an energy and another component could be a position (or a direction). 

Then ``the statistical model'', or simply ``the model'' is
\begin{equation}
p(\boldx| \mu), 
\end{equation}
with the probability or probability density (from context) characterizing everything that determines the probabilities/densities of the observations, from laws of physics to experimental setup and protocol. (This could be more properly be written as $p_X(\boldx| \mu)$; in HEP the subscript is nearly always suppressed, with the function inferred from the argument.) I do not try to use the model notation attributed to Fisher (SIST p.\ 132).

(I use the ``conditional'' vertical bar rather than a semi-colon even in a frequentist context where there is no distribution in $\mu$, as in \citet{kendall1999}. While disfavored in SIST (p.\ 205), I do not find that it ever causes confusion in practice.)  

Thus, a large fraction of what we do in HEP can be characterized as testing the ``point'' null hypothesis,
\begin{equation}
\label{protoa}
H_0\mbox{: }\mu=0.
\end{equation}
vs.\ the continuous alternative, 
\begin{equation}
\label{protob}
H_1\mbox{: }\mu>0.
\end{equation}
I refer to this as the {\em HEP prototype test}.  The more general cases where $\mu<0$, or where the point null is non-zero (as in Section~\ref{higgs}), also exist.  However, much of the foundational discussion can take place with regard to this HEP prototype test.  

A key point is that frequently {\em the model is only defined for $\mu\ge0$}.  For example, if $\mu$ is the mean of a Poisson distribution, or a particle mass, or a reaction rate, then $p(\boldx| \mu)$ {\em does not exist} for $\mu<0$: one would not know how to write a MC simulation of the experimental setup!  On the other hand, there is typically {\em nothing} wrong with (in 1D) $x<0$; it is just a sampled value from the model.  For example, if the model is $p(x| \mu) \sim N(\mu,1)$ with $\mu<<1$, then negative values of $x$ are obtained in nearly half the observations. This has led to a lot of confusion over the years, when $x$ is unfortunately called the ``measured value of $\mu$'', with negative measured values called ``unphysical".  We tried to dispel this confusion in \citet*{feldman1998}, not completely successfully.  In any case, the situation in which one observes $x<0$ for $\mu\ge0$ remains a topic of much discussion~\citep{mandelkern2002,cousinsvirtual,cousins2011}, with multiple methods employed.

For those who might ask if a zero-width ``point'' null value ever truly exists, I address this question in \cite{cousinsJL}.  As has been noted in the statistics literature, all that matters is that any small width of the set of $\mu$ in $H_0$ be much narrower than the width of the likelihood function (i.e., the resolution of the measurement apparatus). Examples are plentiful in HEP.

\subsection{Duality of intervals and tests of nested hypotheses}
The duality between confidence intervals for $\mu$ and frequentist hypothesis tests in the HEP prototype test was implicit in work in our field for many decades, and has been more explicitly advertised by myself and others for over 25 years. That there is {\em not} a duality between Bayesian credible intervals and the traditional Bayesian treatment of this hypothesis test by the method of Jeffreys leads to the Jeffreys--Lindley paradox \citep{cousinsJL}.  Thus, the HEP prototype test of Eqs.~\ref{protoa}--\ref{protob} is a great forum for discussion, with controversies both within the frequentist approach and in comparisons to the usual Bayesian approach (and alternatives).
 
We review here some well-known aspects of the duality of testing and intervals because it seems (from somewhat puzzling discussions of $p$-values versus confidence intervals) that it may not be fully internalized by all practitioners in all sciences. \citet{neyman1937} presented the theory of confidence intervals by deriving them {\em ab initio} using the concept of acceptance intervals and acceptance regions without reference to hypothesis testing. In contrast, it seems that a typical modern treatment first introduces hypothesis testing and then, as already mentioned above, simply defines a confidence interval at confidence level 1$-$$\alpha$ to consist of all values of $\mu$ that are {\em not} rejected by a hypothesis test of size $\alpha$.  (I ignore any distinction between confidence coefficient and confidence level.)

A key aspect of both sides of the duality is that one must specify an {\em ordering} that ranks points in the sample space $\boldx$ \citep[cited by SIST p.\ 131]{pearson1947ordering}. This allows ``extreme" to be defined. {\em I use the notation $\rofx$ (\,$\mathrm{r}$ for rank) for the ordering rule,} with the convention that {\em high rank means not extreme}. The duality of course requires consistent use of the chosen ordering.  

In SIST, an ordering $\dofx$ is defined on p.\ 129 as ``a measure of accordance or discordance, fit, or misfit, \dofx between possible answers (hypotheses) and data''. 
(Evidently, high discordance corresponds to my low rank.) That is fine, but I prefer the more abstract notion of ordering \boldx values, since depending on the context, discordance may have a crucial arithmetic sign, or it might be mapped monotonically to some other function. 

When $x$ and $\mu$ are 1D (so that I remove the boldface), we define the {\em four} different orderings in use in HEP, as follows. If $p(x| \mu)$ is such that higher $\mu$ implies higher average $x$, then we have the three traditional orderings:

\begin{description}
\item{\em \bf Upper limit ordering} Order $x$ from largest $x$ (largest $\rofxNOBOLD$) to smallest $x$: the smallest values of $x$ are the most extreme (and so have the smallest $\rofxNOBOLD$).  Given $\xzero$, the confidence interval that contains $\mu$ for which $\xzero$ is not extreme will normally not contain the largest values of $\mu$.  This leads to confidence intervals known in HEP as {\em upper limits} on $\mu$, which I write here as $\mu<\muUL$ (for values of $\mu$ that exist in the model).  (I believe that some statisticians call them upper confidence {\em bounds}.)
\item{\em \bf Lower limit ordering} Order $x$ from smallest $x$ (largest $\rofxNOBOLD$) to largest $x$, so that the largest values of $x$ are most extreme (and so have the smallest $\rofxNOBOLD$). This leads to {\em lower limits} $\muLL$: $\mu> \muLL$.
\item {\em \bf Central ordering} Order $x$ using {\em central} quantiles of $p($x$| \mu)$, with the quantiles shorter in $x$ (least integrated probability of $x$) containing higher-ranked $x$ (higher $\rofxNOBOLD$); lower-ranked values of $x$ are added as the central quantile gets longer (contains more integrated probability of $x$).  This leads to {\em central} confidence intervals for $\mu$, a special case of more general $[\mulo, \muhi]$.
\end{description}
For any \cl, the endpoints of central confidence intervals with CL are the same as upper/lower limits with confidence level given by $1-(1-\cl)/2 = (1+\cl)/2$.

Both upper limits and central confidence intervals are ubiquitous in HEP. Lower limits are generally only implicitly considered via the duality with hypothesis testing. There are, however, long-standing controversies surrounding these three traditional orderings in the HEP prototype test, since the resulting confidence interval can be the empty set (also described by some as a non-empty interval that is ``nonphysical''). In a related issue, parameter values can be ruled out when there is negligible sensitivity. Historically, this has resulted in various alternatives \citep{cousinsvirtual} to the traditional upper limit ordering, including a Bayesian-inspired method and a popular ``modification'' of the frequentist CL  approach~\citep[Sections 40.4.1 and 40.4.2.5]{readjphysg,pdg2026}.

Furthermore, since these orderings are defined only when \boldx is 1D, a more general ordering principle is needed for multi-D \boldx. In 1998, fellow physicist Gary Feldman and I advocated a {\em unified} ordering based on a likelihood ratio (LR) statistic, which we belatedly learned was the standard LR test statistic, described in ``Kendall and Stuart'' and successors~\citep{kendall1999} for decades:
\begin{description}
\item {\em \bf Likelihood ratio ordering} Order \boldx from high to low values of the likelihood ratio $\Lambda = \lhood(\boldx|\mu)/\lhood(\boldx|\hat\mu)$, where $\hat\mu$ is that value of $\mu$ that maximizes $\lhood$.  Then $\rofx$ is any monotonic order-preserving function of $\Lambda$.
\end{description}
Of course, as SIST discusses on pp.\ 133 ff, a similar ratio was featured in the classic Neyman-Pearson papers, and has been found to be useful even when the strict criteria of the Neyman-Pearson Lemma (simple vs.\ simple) are not satisfied.

As noted above, a key aspect of the HEP prototype test is that the model is only defined for $\mu\ge0$.  As the maximization of $\lhood$ must respect this, it is usual that $\hat\mu=0$ for various \boldx. This can change the unified LR ordering materially from all of the first three orderings.

In the HEP prototype test, this LR ordering ``unifies'' the previously disjoint sets of upper limits and central confidence intervals into one set of two-sided intervals that are no longer central, and {\em never the empty set}. The lower endpoint of the two-sided interval can be zero, in which case one can view the upper endpoint as an alternative value to report, rather than (or in addition to) the traditional upper limit.  The intervals with LR ordering also resolve another issue where one chooses between upper limits and central intervals based on the observed data (thus invalidating frequentist coverage).  (SIST mentions a preference for two-sided intervals on p.\ 358, but in a different context, the so-called two-sided test, where both signs of $\mu$ exist in the model.)

In summary, after observing $\bxzero$ sampled from the model $p(\boldx|\mu)$, the confidence interval $[\mulo, \muhi]$ contains those values of $\mu$ for which $\bxzero$ is {\em not} ``extreme'' at the chosen \cl, {\em given the chosen ordering of \boldx}.  

\section[Post-data p-values and Z-values]{\boldmath Post-data $p$-values and $Z$-values}
\label{pvalues}

 {\em After} data are obtained, the definition used by Fisher and often taught is, ``The $p$-value is the probability under $H_0$ of obtaining \boldx as extreme {\em or more extreme} than the observed $\xzero$.''.  I prefer the numerically identical (if philosophically different) definition,  
``the {\em $p$-value} is the smallest value of $\alpha$ for which $H_0$ would be rejected, {\em had that value been specified in advance}".  Since $\alpha$ is {\em specified in advance} in N-P theory, and since the $p$-value technically has no chance of being the value of $\alpha$ actually specified in advance, my hope is that this definition helps to avoid misinterpretation. It may also help with the interpretation of \pcl, introduced in the next section.
(This is also discussed on SIST p.\ 175, citing~\citet{lehmann}.) 

In HEP, a $p$-value is normally converted to a $Z$-value (unfortunately commonly called ``the significance $S$'' in HEP), which is the equivalent number of Gaussian standard deviations ($\sigma$). E.g., for a one-tailed test in a search for an {\em excess}, a $p$-value${} = 2.87 \times 10^{-7}$ corresponds to $Z = 5$.  Note that Gaussianity (normality) of the test statistic is normally {\em not} assumed when the $p$-value is computed; this conversion to equivalent Gaussian ``number of $\sigma$'' is intended to ease communication.  This needs to be emphasized when communicating outside HEP, as I hear too often statisticians wondering about assumptions of normality, ironically indicating that our conversion is counterproductive in terms of clarity.

I do not attempt to resolve the long, contentious debate about $p$-values.  But I do teach my students that whatever they are, $p$-values are {\em not} the probability that $H_0$ is true!  And that this misinterpretation of $p$-values is unfortunately so common as to be used as an argument against frequentist statistics.  I ask everyone in HEP to please keep in mind:
\begin{itemize}
\item That a $p$-values is calculated {\em assuming that} $H_0$ {\em is true}, so a (non-zero) $p$-value can hardly tell you the probability that $H_0$ is true!
\item That the calculation of the ``probability that $H_0$ is true'' requires prior(s) to invert the conditional probabilities.
\end{itemize}

Regarding loose language about the interpretation of $p$-values, I am a member of the ``$p$-value police'' (SIST p.\ 204) in that I insist that one should always be very clear about what $p$-values {\em are} (easy to say, but not so easy to interpret), and what they {\em are not} (also easy to say correctly, so why be loose and risk encouraging confusion?).

\subsection{Underused vocabulary in statistics?}
\label{missing}
What seems to be missing in most of the statistics literature is a word or phrase in the confidence interval picture that is dual to ``$p$-value'' in the testing picture! I.e., we need a name for that {\em critical value of the confidence level} for which the point-null hypothesis value $\mu_0$ is just on the edge of the {\em post-data} confidence interval or region. 
For this paper, I decided to call it the ``\pcl''. Then clearly, since $\mathrm{\cl} = 1- \alpha$, we have
\begin{equation}
``\pcl\mbox{''} = 1- {}``p\mbox{-value''}.
\label{pcl}
\end{equation}
I.e., having observed \boldx, the \pcl is the smallest value of CL (had it been specified in advance) for which the null hypothesis value of $\mu$ is in the confidence interval. For me, as trivial as this definition is, giving it a name is useful for comparing and contrasting with severity.

An AI tool built by Andrew Brilliant brought to my attention a paper by \citet*{efron1998} that suggests ``attained confidence level" as terminology for 1 minus the $p$-value.  So, the concept of $p$-CL is not new, though it seems to have had little use in the sense of Efron and Tibshirani, while ``attained confidence level" is also used to mean empirical or simulated coverage probability of approximate intervals generated by various methods. There is also a generalization called ``observed confidence level" advocated by Alan Polansky.  In this paper, I stick to $p$-CL, as defined by Eq.~\ref{pcl}.

One can use the old concept of {\em confidence distribution} as a {\em technical device} for reading off the \pcl.  (I have not tried to understand the potential of confidence distributions as claimed by Singh and others, as mentioned on SIST p.\ 391.) In 1D, the (post-data) confidence distribution \citep{cox1958} is the set of all confidence intervals at different \cl.  (See also SIST p.\ 195.) I.e., one can make the post-data plot of the endpoints of the confidence interval as a function of \cl (or provide the equivalent information, as in the likelihood scan in the third example of Section~\ref{higgs}).  Given $\mu_0$, one can then simply read off \pcl from the confidence distribution, i.e., find that \cl for which $\mu_0$ is an endpoint.  This is purely a technical comment, quite apart from any philosophical interpretations that confidence distributions may have.

As \citet{cox1958} noted, ``In applications it will often be enough to specify the confidence distribution, by for example a pair of intervals, and this corresponds to the common practice of quoting say both the 95 per cent and the 99 per cent confidence intervals.'' Examples from HEP providing confidence intervals for a couple CL's are in Section~\ref{hepexamp}.  We can have more complicated examples where the confidence interval or region is not simply connected (as in Section~\ref{sec:sterile}), due to non-linearities (such as sine functions) in the model.  However, the concept of confidence distribution as a technical device for determining \pcl is still workable, even if awkward to display.

I find it surprising that a quantity like \pcl is not commonly used, especially since confidence distributions are well-known.  The result seems to be that confidence intervals are thought to be as rigid as $\alpha$, even post-data, and that there are even claims that scientists view all values of $\mu$ inside a confidence interval as equivalent. They are easily distinguished by using \pcl, which I think comes naturally to physicists in HEP.

\subsection{\boldmath Example of \pcl from HEP}

Most papers in HEP use conventional values of CL, so that one needs to use the information given to infer a precise value of \pcl for some $\mu$ of particular interest.  A memorable example comes from early attempts to measure an abstract angular quantity known as $\sin2\beta$, deeply related to differences between matter and antimatter in the equations of physics.  The statistical model $p(\boldx|\sin2\beta)$ is only defined for ${-}1\le\sin2\beta\le1$ (but for {\em all} \boldx). The bounded parameter space made this a natural application of the LR ordering of \boldx for obtaining two-sided confidence intervals in $\sin2\beta$.  For the CDF collaboration's first attempt at Fermilab~\citep{cdffc}, \boldx was a sample from a normal distribution centered on the unknown true value of $\sin2\beta$ with $\sigma=0.42$. The obtained value was $\xzero = 0.79$.  As $\sin2\beta=0$ is a special value in the physics equations (matter--antimatter symmetry), they calculated that value of CL for which 0 was an endpoint of the confidence interval, and obtained 93\% for what I call \pcl.  This of course corresponds to a $p$-value of 7\% in the LR ordering.  (Later experiments were much more precise.)

\subsection{Power is (mostly) a pre-data concept in HEP}

Since this paper focuses on {\em post-data} practices in HEP, I only briefly mention our general {\em lack} of use of the {\em pre-data} concept of power when we perform {\em post-data} inference and hypothesis testing. Power is crucial for understanding projected sensitivity as a function of sample size, choice of test statistic, comparison of experiments, etc. In searches for newly observed particles, it is common to characterize an experiment by providing the {\em median} $p$-value (or $Z$-value) that is expected for some putative signal rate (for an assumed sample size). This is equivalent to providing $\alpha$ for $\beta = 0.5$ (and hence power of $0.5$). Median pre-data sensitivity is generally reported along with the experimental results, nowadays often along with a couple quantiles of expected sensitivity. This is a form of model-checking that can help to interpret a situation where one experiment has a shorter post-data confidence interval than another, i.e., whether this was expected from pre-data power considerations, or a consequence of statistical fluctuations in the data. 

A contentious case arises when an upper limit (UL) excludes values of a parameter for which there was very little sensitivity to detect a possible signal.  For example, consider the HEP prototype test of Section~\ref{proto} with a model $p(x| \mu) \sim N(\mu,1)$ with $\mu\ge0$ (say a measurement of neutrino mass), and suppose that the sampled value is, say, $x=-1.9$ or $x=-2.1$.  The null hypothesis of $\mu=0$ will not be rejected in favor of $\mu>0$, and one desires to place a 95\% CL UL on $\mu$.  Depending on the method, the UL will be the empty set, a non-physical value of $\mu$, or a positive number much smaller than the rms deviation of the measurement (resolution of the measuring apparatus), here 1 unit.  This is considered unsatisfactory to many in HEP, and various ad hoc patches exist.  One's attitude toward these methods depends on one's speculation about the root problem (if one exists), but it includes model mis-specification (unknown systematic effects) or mis-measurement.  My virtual talk \citep{cousinsvirtual} gives a brief history and enumeration of the various methods, including one case that is an explicit use of pre-data power to constrain the post-data calculation of the UL~\citep{pcl}. I have opposed its use, but it is used in some communities (e.g., searches for direct detection of dark matter). I tend to view such ad hoc methods as protection against model mis-specification, and that the best thing is to supply consumers with a completely transparent explanation of the situation and all the cross-checks performed, so that readers can make up their own mind regarding an appropriate inference.

Thus, apart from this contentious case of upper limits near a parameter boundary, there is little use of power in {\em post-data inference} in HEP. I was surprised to learn from SIST and from \citet{HoenigHeisey2001} (cited by SIST on p.\ 357) about controversies related to post-data use of power and some variants. The controversial recommendations about using post-data power analysis (panned by SIST and by Hoenig and Heisey) do not exist in usual HEP practice, as far as I know. Similarly, Stephen~\citet[Section 13.2.13]  {sennbook2021} criticizes the notion that, ``In order to interpret a trial it is necessary to know its power", noting that ``This is a rather silly point of view that nevertheless continues to attract adherents. A power calculation is used for planning trials and is effectively superseded once the data are in."  Since common practice in HEP is aligned with this view, in the remainder of this paper regarding post-data inference, I do not mention use of power in HEP.

\section{Sketch of top-level data analysis procedure in HEP}

In a search corresponding to the HEP prototype test, the $p$-value for the null hypothesis will normally be reported (often via the corresponding $Z$-value).  In the interval picture, this is (as in Section~\ref{missing}), $1-\pcl$ for the {\em lower limit} for $\mu$, though I think that practitioners rarely think about it that way: for testing the null hypothesis, the testing picture mentality is typical.

If the $p$-value is small enough to indicate the possibility of a non-zero value of $\mu$ in nature, then generally a {\it central} confidence interval is also reported.  To statisticians, this is the so-called {\em effect size} in original units \citep{cousinsJL}, much discussed in social science and medical literature.  As described in SIST (p.\ 211), this can then be used to test a theoretical prediction, if one exists (as was the case for the Higgs boson). 

I recall some discussions in HEP about potential bias in confidence intervals because of post-data selection of models. However, typically there is no attempt to correct for any such bias if the null hypothesis ($\mu=0$) is rejected and a confidence interval for $\mu$ is reported. (The {\em unified approach} of \citet{feldman1998} is however commonly used in some subfields of HEP and mitigates a similar bias.)

If the $p$-value is not small (and sometimes even if it is), {\em upper} limit(s) for $\mu$ under the alternative hypothesis $H_1$ are also reported, for some conventional confidence level(s) (90\% in some subfields of HEP, 95\% at the LHC).  Many papers also convey enough information so that the upper limit at other confidence levels can be approximately inferred. As mentioned in Section~\ref{background}, it is common to replace the strict frequentist upper limits with one of the alternatives mentioned. But this does not affect the conceptual comparison to severity.

Thus, when reporting these three results of the analysis, the {\em ordering} (ranking) of \boldx in each is different (in fact opposite for upper and lower limits).  In (very common) cases where upper limits are shown, we strongly discourage the practice of saying, (paraphrasing) ``We have no evidence for a signal [for $\mu>0$], and so we calculate an upper limit''.  As mentioned above, \citet*{feldman1998} criticized such a procedure, which we dubbed ``flip-flopping'' (using the {\em observed data} to choose the ordering for testing/intervals).  We advocated the {\em unified} LR ordering as mentioned, a single ordering replacing all three orderings. The LR ordering can of course be consistently used for both intervals and testing, per the duality, for a variety of CL's.

It is easy to show that the $p$-value obtained for testing $H_0$ is the same whether one uses LR ordering or the traditional lower-limit ordering. I.e., in the interval picture, given $\mu_0=0$, the \pcl for lower limits is the same as that for unified LR intervals.  For tests of other $\mu$, the \pcl's and upper limits from the two orderings are different.  In certain subfields of HEP (for example neutrino physics), results using the unified LR ordering are routinely included in reported experimental results, even if the traditional orderings are quoted as well.

From an operational point of view (i.e., in computer code), in the first three traditional orderings of Section~\ref{background}, the identification of the null and alternative hypotheses changes from ordering to ordering. Physicists at the LHC had to understand this in order to write general-purpose software with a guide to its use.  In the HEP prototype test, for computing the $p$-value, the software treats $\mu=0$ as the null hypothesis, as discussed.  For computing the upper limit, the software in effect steps through the various positive values of $\mu$ one at a time, and for each value considers it to be a null hypothesis for a one-sided test in the opposite direction as that used to test $\mu=0$.

Thus, I was interested to see that SIST (p.\ 346) cites Stephen \cite{sennbook2021} as saying something similar: ``Senn gives another way to view the severity assessment of $\delta>\delta^\prime$, namely `adopt [$\delta=\delta^\prime$] as a null hypothesis and then turn the significance test machinery on it' ''.  I did not quite follow why SIST then says that, ``the error statistician is loath to advocate modifying the null hypothesis'', and in the next section, ``We are not changing the original null and alternative hypotheses!''. These may be philosophical statements, but operationally the calculation (computer code) is as if that change in null hypothesis is being made.

The argument seems to hinge on whether one is ``allowed'' to switch from lower limits to upper limits.  In HEP, the usual procedure certainly does so.  (Lower limits are nearly always considered implicitly via the dual testing picture, while upper limits are conveyed in the interval picture.)

Whether speaking of two-sided intervals or upper limits, in HEP we tend to talk about ``exclusion'' of values of $\mu$ that are outside the confidence interval, at the quoted CL. E.g., one might say, ``Values of $\mu$ above $\muUL$ are excluded at 95\% C.L.''

In hopes that practitioners of severity can compare more easily to practice in HEP, some defining equations can be written. Let us focus on the interval picture, and let $\rofx$ represent one of the four {\em orderings} for \boldx described in Section~\ref{background}. Post-data, after observing $\xzero$ and not rejecting $H_0$, the statement that ``$\mu<\muUL$ at 95\% \cl'' means that for $\mu>\muUL$, $\xzero$ is {\em not} in the highest-ranked 95\% values of \boldx using the ordering corresponding to upper limits.  That is,
\begin{eqnarray}
\label{ularray1}
0.95 & = & \PR(\rofx > \rxzero; \mu=\muUL) \nonumber \\
     & \le & \PR(\rofx > \rxzero; \mu>\muUL).
\end{eqnarray}
For some $\mu_1$ of scientific interest that is not equal to $\muUL$, one can find its \pcl for this ordering by finding $\mu_1$ (perhaps iteratively) such that,
\begin{equation}
\label{pclmu1}
\pcl = \PR(\rofx > \rxzero; \mu=\mu_1).
\end{equation}

If $H_0$ is rejected (and even if not), in HEP one will usually quote, at one or more conventional \cl, a two-sided confidence interval $[\mulo, \muhi]$ obtained from either central ordering or LR ordering.  This means, for that ordering, that $\xzero$ is in the CL top-ranked values of \boldx for $\mu$ in the confidence interval, and that $\xzero$ is in lower-ranked values of \boldx for $\mu$ outside the confidence interval:
\begin{eqnarray}
\label{centralarray1}
\cl & \ge & \PR(\rofx \le \rxzero; \mu\in [\mulo, \muhi]) \mbox{~and} \\
\cl & <   & \PR(\rofx > \rxzero; \mu\notin [\mulo, \muhi]).
\end{eqnarray}

For some $\mu_1$ of scientific interest that is not one of the endpoints of a quoted conventional confidence interval, one can find its \pcl, that value of CL for which $\mu_1$ is just on the edge of the interval, either $\mu_1=\mulo$ or $\mu_1=\muhi$, with
\begin{equation}
\label{centralarray2}
\pcl = \PR(\rofx > \rxzero;  \mu\in [\mulo, \muhi]).
\end{equation}

\subsection{Statistician's description of Higgs boson discovery}
Statistician David van Dyk (a self-described Bayesian who cares about frequentist performance) studied the methods that the CMS and ATLAS collaborations had used in the search for the Higgs boson, drawing on extensive discussions with the physicists who developed the analysis methods.  He reported his findings in \citet{vandyk2014}, using a simplified model for illustration.  
 
The more complicated methods actually used, with multitudes of nuisance parameters and combinations of different production and decay mechanisms of the Higgs boson, are documented in: \citet*{CMS-NOTE-2011-005} (written in 2011 in anticipation of forthcoming data); \citet{cms_limits_2012} (limit setting in 2012 with a preliminary excess); and \citet{atlashiggs,cmshiggs} (observation papers with brief overview of statistics methods and description of physics processes and observations).  Discussion of the more recent software tools in the CMS collaboration is in \citet{combine}.

\section{Examples from HEP}
\label{hepexamp}
Before getting to the prototype HEP test, I begin with an example where only two values of the parameter of interest exist in the model (thanks to quantization of quantum mechanics).  Then we turn to examples from neutrino physics and measurements of properties of the Higgs boson. In plots of upper limits or the multi-D equivalent in HEP, often only one \cl is chosen; the rest of the confidence distribution must be inferred by the reader, as I think experienced readers generally do. However, some authors consider it of interest to plot the upper limit for at least one other value of CL, as mentioned in the quote by Cox in Section~\ref{missing}.  

\subsection{Simple versus simple hypotheses}

I do not know of any cases in HEP where both $H_0$ and $H_1$ are completely ``simple" in the language of Neyman and Pearson, with no unfixed parameters. However, in addition to binary classification problems already mentioned, we do have prominent examples where the parameter of interest is fixed in both hypotheses, and where the presence of unfixed nuisance parameters does not badly spoil the ``simplicity''.  For example, the hypothesis test of $H_0$ vs.\ $H_1$ might be:
\begin{itemize}
\item ``1 unit of spin angular momentum" vs.\ ``2 units of spin angular momentum" for a new heavy resonance that decays to two particles; or
\item ``$J^P=0^-$" vs.\ ``$J^P=0^+$" for the Higgs-like boson (symbols explained below); or
\item ``this particular muon has positive electric charge $+$1" vs.\ ``this particular muon has negative electric charge $-$1". (The magnitude of a muon's electric charge is always $+$1 in the usual units. The sign is determined from the curvature in a magnetic field, and can be mistaken at very high momentum, when the trajectory is nearly straight.)
\end{itemize}
In the first two examples, the discreteness and separation of the two hypotheses is a feature of quantum mechanics, which is so-named because the allowed values of observables such as these are {\em quantized}. The quantization of electric charge is a longer story, but a muon cannot have an electric charge between $-$1 and $+$1. (Quarks and antiquarks do have charges of $\pm$1/3 and $\pm$2/3, but not continuous values.)

Thus, I was surprised to see that \citet*[p.\ 176]{mayospanos2011error} state that ``\dots in a proper test, the null and alternative hypotheses must exhaust
the parameter space, and thus, ``point-against-point” hypotheses are at best highly artificial, at worst, illegitimate." Having pointed to quantization in quantum mechanics, I hope we can agree that there are examples that are neither artificial nor illegitimate. Framing tests such as these in this ``simple" way does make strong assumptions, in particular that one of the two hypotheses is true. As discussed below, it is good practice to check these assumptions when data are in hand, just like any other situation with goodness-of-fit tests or model-checking. In HEP, we routinely check the validity of model assumptions (and there might be a Nobel Prize for someone if they fail). 

A simplified illustration of the spin angular momentum test is in Appendix~A of my lectures \citep{cousins-slides}, which summarizes \citet{cousinsspin}. Here I discuss the example of an early $J^P$ test published by the CMS Collaboration. 

\subsubsection[Early Higgs boson spin-parity test of 0- 
vs.\ 0+]{\boldmath Early Higgs boson spin-parity test 
of $0^-$ vs.\ $0^+$}
\label{higgsCP}

Elementary particles have various properties, referred to as intrinsic ``quantum numbers", that take on only discrete values, as the name suggests.  These include intrinsic spin angular momentum (denoted by $S$ or $J$), which can have either integer or half-odd-integer values. Another is the intrinsic parity $P$, which describes the change (+1 or ${-}1$) in the sign of the state vector when the coordinate system is reflected through the origin. The theory of the Higgs boson specifies that it has $J=0$ and $P=+1$, written as  $J^P=0^+$ (in a word, ``scalar"). 

There exist many other particles with $J=0$ (including the ubiquitous pion), but they all have $P=-1$, i.e., $J^P=0^-$ (in a word, ``pseudoscalar").  Thus, when the new boson was discovered in 2012, it was crucial to confirm (or not) that it had positive $P$. (Inferences about $J$ were also important, but here we focus on $P$.) In the theory, the angular distributions of the products of the Higgs boson decays carry spin and parity information. With the vector $\boldx$ now denoting the energy and momentum measurements of each decay, there were pdfs $p(\boldx; J^P{=}0^+$) and $p(\boldx; J^P{=}0^-$).  (Here I follow the physicist's convention of using different arguments to indicate implicitly different pdfs, rather than subscripts on $p$.) 

Inserting the observed $\boldx$ from each Higgs boson decay into the pdfs provides one value of each of the likelihoods $\lhood(J^P{=}0^+)$ and $\lhood(J^P{=}0^-)$.  Multiplying all the individual decay likelihoods yields likelihoods (called $\lhood_{0^-}$ and $\lhood_{0^+}$) for each hypothesis, based on the whole data set. We thus obtain from the full data set one value of the discrimination test statistic, which is $-2\ln (\lhood_{0^-} /\lhood_{0^+})$. Here we implicitly (if arbitrarily) take $J^P=0^-$ to be the null hypothesis, following usual conventions. Nuisance parameters are separately optimized for each $P$ (in the spirit of profile likelihood).

\begin{figure}
\begin{center}
\includegraphics[width=0.6\textwidth]{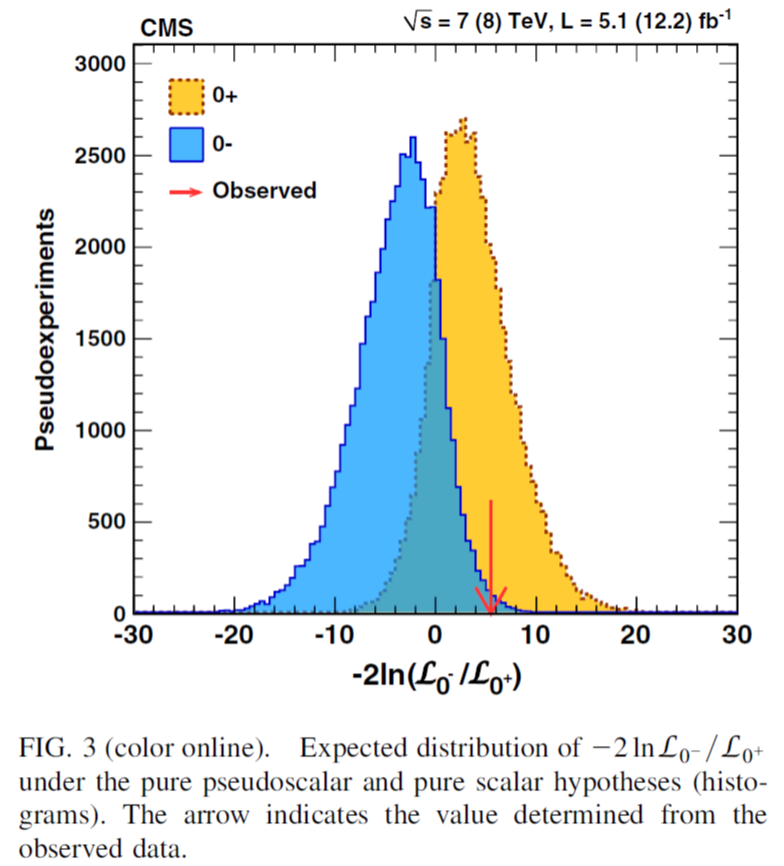}
\caption{Figure and caption from the early CMS paper on a test of   $J^P=0^+$ vs.\ $J^P=0^-$ for the Higgs-like   boson~\citep{cmshiggsprop2012}.}
\label{cmsjp}
\end{center}
\end{figure}

An early result from the CMS collaboration \citep{cmshiggsprop2012} is shown in Fig.~\ref{cmsjp}.  The pdf of the test statistic, obtained by MC simulation, is shown for each of the two hypotheses. The large sum of logs in the test statistic leads to the evident good approximation to a normal distribution in each case, even though the underlying pdfs are far from normal. (See also the spin discrimination papers mentioned above for striking examples of the central limit theorem in action.) The single observed value of the test statistic in this data set is indicated by the red arrow at 5.5.

As this was an early publication by the CMS Collaboration that approximated a simple-versus-simple test, there was much discussion within the collaboration about how best to present the results post-data. E.g., the labeling of the hypotheses as $H_0$ and $H_1$ is rather arbitrary. CMS reported:
\begin{enumerate}
\item The observed value $-2\ln(\lhood_{0^-} /\lhood_{0^+}) = 5.5$, favoring $0^+$; 
\item For the pseudoscalar hypothesis $H_0$: $J^P = 0^-$,  the $p$-value $= 0.0072$;
\item For the scalar hypothesis $H_1$: $J^P = 0^+$, the $p$-value $= 0.7$;
\item Finally, \cls${} = (0.0072)/(1-0.7) = 0.024$, ``a more conservative value for judging whether the observed data are compatible with $0^-$.'' (This is an invention in HEP~\cite[Section 40.4.2.5]{readjphysg,pdg2026}.)
\end{enumerate} For each $p$-value calculation, the relevant tail probability is for the tail in the direction of the other hypothesis.

Although we did not explicitly report it, the Bayes factor for this test is closely related to the observed value of the reported likelihood ratio, as its derivation differs only in the treatment of the nuisance parameters (marginalization rather than separate optimization).

I do not recall that CMS had any official pre-data critical value of Type I error probability $\alpha$ in mind; in a huge collaboration, opinions can differ, and there is no convention in this case. The paper concluded, ``Under the assumption of spin zero, the observed data are consistent with the pure scalar hypothesis, while disfavoring the pure pseudoscalar hypothesis."  Further investigation by the ATLAS and CMS collaborations established the consensus that $J^P = 0^+$, and the ``Higgs-like" boson was subsequently referred to as ``{\em a} Higgs boson" (leaving open the possibility there might be another, as widely speculated). As was also much discussed, the strong assumption of pure $0^+$ or pure $0^-$ as the only allowed possibilities can be relaxed in speculative models. One then investigates the possibility that the Higgs boson, while predominantly $0^+$,  has a quantum-mechanical admixture of $0^-$, with a continuous mixing parameter being the parameter of interest~\citep{cpmixtureHIG-25-012}. 

I find it interesting that Bill Thompson~\citep[p.\ 108]{thompson2007} argues that the set of the {\em two} $p$-values is the irreducible post-data ``evidence'' in a situation like this, and many in HEP may agree. Certainly neglecting one of the $p$-values can be dangerous. As an example from \citet{cousinsJL}, for large sample size, it is conceivable that $H_0$ is rejected at 5$\sigma$, while if $H_1$ were the null hypothesis, it would be rejected at 7$\sigma$.  A physicist would be well aware of this circumstance and not fall into the straw-man trap of implicitly accepting $H_1$ by focusing only on the $p$-value for $H_0$ and ``rejecting'' (only) $H_0$.  The natural reaction would be to question both hypotheses and launch renewed investigations of model uncertainty, other systematic uncertainty, and (of course) bugs throughout.

More tests of the compatibility of the particle's properties with ``the Higgs boson" are described in Section~\ref{higgs}.

\subsubsection[{Severity discussion of the Higgs boson JP analysis}]{Severity discussion of the Higgs boson $J^P$ analysis}

Taking first the case with $H_0$: $J^P = 0^-$, the $p$-value is 0.0072. While CMS was characteristically cautious in merely concluding that $H_0$ was ``disfavored", for the discussion here let us use a convention from another community and say that $H_0$ is ``rejected", placing us in the domain of ``Severity Interpretation of Rejection" (SIR). That is, the initial null hypothesis is $H_0$, so it is the initial claim. Once $H_0$ is rejected, the new claim $C$ is ``$H_0$ is false". We need to apply severity to that claim. I find that applying the discussion regarding SIR throughout SIST to be somewhat confusing in this case, even after consulting pages 143, 265-6, and 351. So I revert to the criteria of (S-1) and (S-2) on p.\ 92 of SIST, quoted above in the Introduction.

Proceeding with these criteria, upon rejection of $H_0$, $H$ is the claim, ``$H_0$ is false".  
Then S-1 demands that the claim agrees with the data. 
Given ``the strong assumptions, in particular that one of the two hypotheses is true", mentioned above, then one could say that the claim ``$H_0$ is false" is the same claim as ``$H_1$ is true".  Whether this agrees with the data is tested by the second $p$-value (item 3 above), which is 0.7, certainly not in disagreement with the data. So I would say that criterion S-1 is passed. 

Meanwhile S-2 reads, ``with high probability, `$H_0$ is false' would not have passed the test so well, were $H_0$ true." 
Given the low $p$-value 0.0072 for $H_0$ and the high $p$-value for $H_1$ in items 2 and 3 above, I would say that criterion S-2 is passed. Thus, the rejection of ``$H_0$ passes both severity checks.

As a check, I can also apply the rephrasing on p.\ 143 that, after rejection of $H_0$, ``severity requires that there be at least a reasonable probability of a worse fit with $C$, if $C$ is false (S-2)".  But when C is ``$H_0$ is false", ``$C$ is false" again means ``$H_0$ is true". And as before, under our strong assumptions, ``$H_0$ is false" means ``$H_1$ is true", So severity requires that there be at least a reasonable probability of a worse fit with ``$H_1$ is true", if ``$H_0$ is true". Looking at the red arrow and the two distributions in Fig.~\ref{cmsjp}, this is the case.

Thus, from the point of view of a practicing experimental physicist, it seems that the severity criteria uses the same two $p$-values that a physicist would use, in a very similar manner, with just differences in nomenclature. (I leave any philosophical distinction to others.)

Looking elsewhere in the writings of Mayo and Spanos fir similar situations, 
their 2011 review mentioned above \citep*[p.\ 175]{mayospanos2011error} gives a worked example to refute the ``well-known canard", ``Whether there is a statistically significant difference from the null depends on which is the null and which is the alternative." I am not sure where this canard comes from, but I agree with the (apparent) conclusion that one can test each of the two hypotheses individually for discrepancy in the direction of the other hypothesis, and that rejecting one of them does not automatically imply that the other is true; one must test the other one.  It seems to me that the way that we view this in HEP using the language of $p$-values reaches the same scientific conclusions as would Mayo and Spanos, but I am not aware of a ``canard" that needs to be addressed. I also do not see a need to substitute the severity concept in order to reach the correct scientific conclusions. 

A more complete discussion would consider model adequacy and the various possibilities. If we imagine Fig.~\ref{cmsjp} with much larger sample size, so that the distributions are much farther apart (with respect to their widths), then the observed test statistic could be: 
\begin{itemize}
    \item so far to the left (or right) that it casts doubt on both hypotheses; 
    \item just to the left or right of one of the hypotheses, and hence deemed compatible with it, while the other $p$-value indicates incompatibility with the other hypothesis; 
    \item in between the distributions for the two hypotheses and apparently incompatible with both hypotheses. 
    \end{itemize}
    It is hard for me to imagine that a physicist in HEP would consider only one of the $p$-values and commit one of the fallacies.  Of course, the inferential judgement will depend on the degree of surprise, and many cross-checks would be performed. Meanwhile, some physicists may prefer focusing on the likelihood ratio. For an enlightening comparison of $p$-values and the likelihood ratio, see \citet{demortierlyons}.

\subsubsection{\texorpdfstring{A large \textbf{\textit{n}} problem?}{A large n problem?}}
From SIST (pp.\ 240 ff) and \citet{spanos2014} and other contexts, 
I have learned that there is considered to be a ``large $n$ problem" in some circles.  Meanwhile, in HEP, considerable resources are devoted to obtaining as large a sample size as is feasible.  The LHC is currently shut down for expensive (well over a billion Swiss francs) upgrades to the accelerator and the detectors in order to increase the sample sizes by at least factor of 10 in the next 15 years or so.

As $n$ increases, there are technical challenges such as handling enormous data sets, reconstructing overlapping collisions in the detector, etc. Finer systematic effects and new sources of background can be uncovered. 
Furthermore, statistical uncertainties can become smaller than the systematic uncertainties (including model uncertainty in the theoretical calculations).  We are usually well aware of this, since it is routine to report separately the dominant sources of uncertainty.  Sooner or later, increasing $n$ implies yet more effort to reduce the size of systematic uncertainties via subsidiary measurements, improved theory calculations, etc. Ultimately, systematic uncertainties can limit the attainable precision of some measurements.  

Among the systematic effects for which there is an ongoing decades-long world-wide effort to reduce uncertainties are: detailed description of the momenta of quarks and gluons inside colliding protons (``parton distribution functions"); precise calculations in quantum field theory for the production and decay of Higgs bosons and other particles at particle colliders; background shape modeling when the simulations are not reliable; and efficient simulations of the whole chain of collisions, detector responses, and reconstruction of events.

For those concerned about a result that is ``statistically significant" but not ``substantively significant", I invite them to read Section 8 in \citet{cousinsJL}, in particular ``No effect size is too small in core physics models". 

All these challenges with increasing sample size might be characterized as ``problems" or ``challenges" to address, but I would claim that there is no ``large $n$ problem" that induces fallacies for a competent researcher or that casts doubt on the use of $p$-values and confidence intervals (and confidence distributions) in HEP.     

\subsection{A search for sterile neutrinos}
\label{sec:sterile}

The next example concerns a search for evidence of a so-called
``sterile neutrino''.  There are three known neutrino species in the
SM; all interact with other particles via the ``weak force'', one of
the four known forces in nature.  There is long-standing speculation
about the existence of (at least) a fourth neutrino, called
``sterile'' because it does not interact with the known weak
force. One way to detect the presence of a sterile neutrino is if one
of the ordinary neutrinos transmutes into another neutrino via an
interaction with a sterile neutrino mediated by a new, even weaker
force of nature.

In a simple case, the probability of transmutation can be expressed in terms of two unknown parameters in the speculative models, namely the difference in mass-squared, $\Delta m^2$, between the sterile neutrino and another neutrino; and an effective angle $\theta$ in a quantum-mechanical Hilbert space.  Experiments normally report 2D confidence regions in a plane conveying the two parameters, usually plotted as $\Delta m^2$ vs.\ $\sin^22\theta$.  So for this example, we consider $\mu$ to be the two-component vector with these two parameters, and the null hypothesis of Eq.~\ref{protoa} means $\Delta m^2 = \sin^22\theta = 0$.  (In fact, both parameters must be non-zero for the transmutation to take place, but the resulting experimental degeneracy along two axes is not relevant here.)

Experiments named LSND and MiniBooNE have claimed evidence for such transmutations, and have hence reported confidence regions that exclude $H_0$.  Numerous other experiments have searched for such transmutations and not seen them, i.e., they have reported observations compatible with $H_0$.  The plot I chose as a case study is one from a 2013 paper by the ICARUS collaboration \citep{Segreto:2013esa}, which obtained results compatible with no transmutation, but the results were {\em not} stringent enough to rule out the evidence from the previous LSND and MiniBooNE experiments. 

\begin{figure}[t]
\centering
\includegraphics[width=0.7\textwidth]{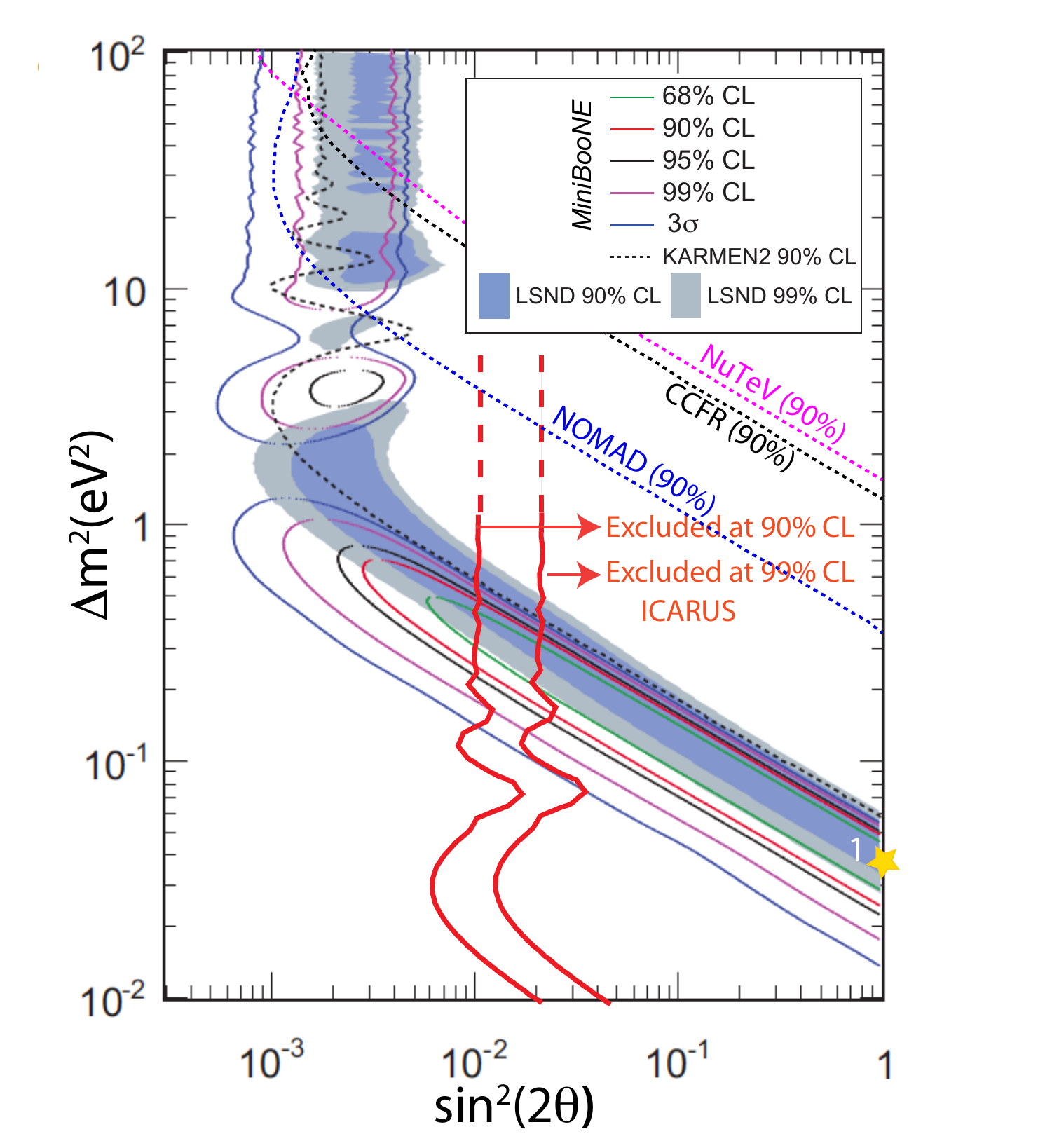}
\caption{Plot of confidence regions in the plane of $\Delta m^2$   vs.\ $\sin^22\theta$ for sterile neutrino searches.  The ICARUS   2013 results (regions to the left of the red curves), do not reject the null hypothesis (the origin); the CL labeled on the ``exclusion'' curves corresponds to the claims that the true parameters are to the left of the respective curve.  Also shown are the confidence regions of LSND and MiniBooNE that reject the null hypothesis (!); the values of CL given in the legend correspond to the interior of the regions outlined by the various curves. There are also curves from some other less sensitive experiments. From  \protect\citet{Segreto:2013esa}, which gives references for all curves.}
\label{sterile}
\end{figure}

In Figure~\ref{sterile}, the 2D confidence regions from ICARUS data are to the left of the red vertical curves, shown for 90\% and 99\% \cl.  They are compatible with the null hypothesis, $\Delta m^2 =
\sin^22\theta = 0$.  

The plot also shows the confidence regions from LSND and MiniBooNE, which do not include the null hypothesis.  The MiniBooNE regions are given at more than one CL.  The conclusion from the plot is that ICARUS data was compatible with the  \boldx null hypothesis, but it did not have the sensitivity to exclude the parameter values favored by LSND and MiniBooNE.  This was also the case for other experimental curves in the upper left of the figure, and for additional ICARUS data.  

I chose an obsolete plot to make the statistics point, but this subfield of neutrino physics is an exciting area that continues in earnest. At present, the results from the MicroBooNE Collaboration \citep{microboone2025} appear to rule out the sterile neutrino that was first indicated by the LSND collaboration in 1995.

\subsection{Searches for non-standard couplings of the Higgs boson}
\label{higgs}
After the discovery in July 2012 of a particle dubbed a ``Higgs-like boson'', a comprehensive program was undertaken to compare measurements of its properties with those of the Higgs boson of the SM.  In less than a year, enough was learned to remove the qualifier ``-like'' from public discourse, but a decades-long campaign is still considered crucial for testing alternative possibilities.  This campaign includes measuring, as precisely as possible via production and decays, the strength of the interactions of each particle in the SM with the Higgs boson, which can be loosely viewed as the carrier of its own special force.  These interaction and decay strengths are functions of parameters called ``couplings'' in the fundamental equations of the SM. The couplings are predicted with small uncertainty by the SM, incorporating previous precise measurements of other quantities, in particular the particles' masses.  Any confirmed departure of couplings from the SM values would be an exciting observation of ``new physics''.

The results of numerous analyses of experimental data can be summarized by displaying measurements of relative strengths denoted by $\mu$, normalized so that $\mu=1$ is the SM prediction.  As discussed in SIST (p.\ 211) and in \citet[Section 5.4]{cousinsJL}, historically the first thing to establish was that $\mu>0$ for at least some of the couplings, i.e., that there was a new boson that coupled to known particles. Then attention quickly turned away from testing $\mu=0$, and toward testing the SM value $\mu=1$ for as many couplings as possible.  Thus, in this subsection, we depart from the HEP prototype test of Eqs.~\ref{protoa} and \ref{protob}, and instead perform a two-sided test of the point null hypothesis of the SM, 
\begin{equation}
\label{SMa}
H_0\mbox{: }\mu=1,
\end{equation}
vs.\ the continuous alternative 
\begin{equation}
\label{SMb}
H_1\mbox{: }\mu\ne1.
\end{equation}
In speculative physics models beyond the SM, $\mu<1$ and $\mu>1$ are possible.

The three plots shown here are taken from a paper \citep*{atlascmscouplings}, where the two collaborations combined their data samples after first performing their own measurements separately.  The combination was a huge effort taking over a year, going well beyond simply computing a weighted average of the separate measurements, and carefully evaluating the level of correlations among all the nuisance parameters.

\begin{figure}[t]
\begin{center}
\includegraphics[width=0.95\textwidth]{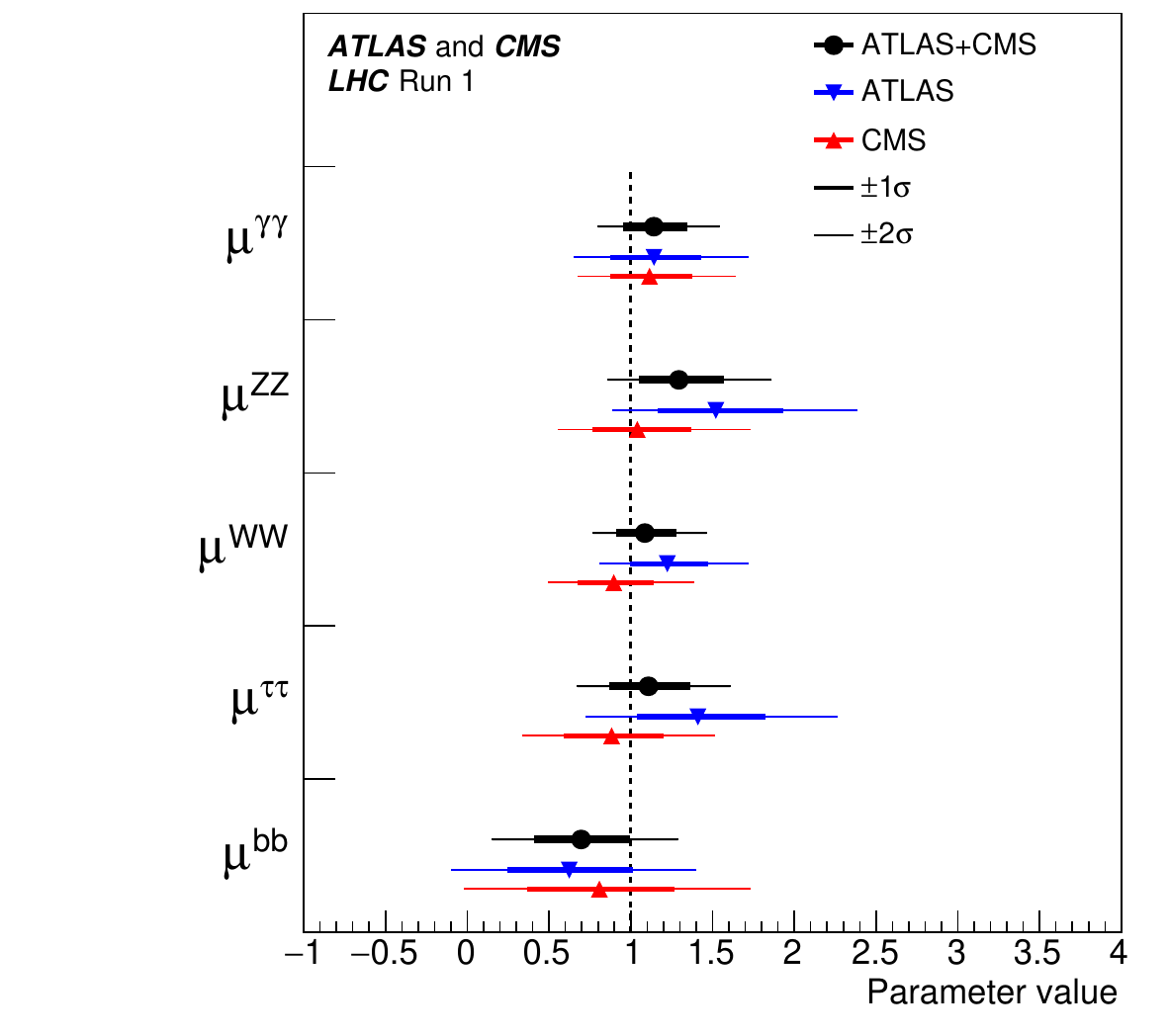}
\end{center}
\caption{Point and interval estimates for the Higgs boson decay strengths from the combination of ATLAS and CMS data in 2016.  Also shown are the results from each experiment.  The error bars indicate the $1\sigma$~(thick lines) and $2\sigma$~(thin lines) confidence intervals. Figure 13 of \citet{atlascmscouplings}.}
\label{fig:muDecay}
\end{figure}

Figure~\ref{fig:muDecay} presents the measurements of the interaction strength as measured in the decay of the Higgs boson to {\em five} different decay modes, each to a pair of known particles: two photons ($\gamma\gamma$); two $Z$ bosons (electrically neutral bosons that are carriers of the weak force); two $W$ bosons (electrically charged bosons that are carriers of the weak force); two tau leptons (particles similar to the electron, but with nearly twice the mass of a proton); and two $b$ quarks (the fifth quark, called ``bottom'' or ``beauty''). The two confidence intervals plotted for each measurement (obtained via Wilks's theorem after appropriate cross-checks) are approximately 68\% and 95\% \cl central intervals, and are part of the confidence distribution of intervals at all \cl.

These measurements from these early data sets were compatible with $\mu=1$, with interval lengths that were quite impressive experimental achievements at that moment, so soon after the discovery.  

\subsubsection{Severity discussion for interaction strengths}

We would say that $\mu$ values displayed are within 10-20\% of unity, with more precise statements available using $p$-CL and the confidence distributions. If one of the measurements is noticeably discrepant in one direction, that would catch our eye and probably motivate theorists to consider speculative explanations for that sign (and magnitude) of discrepancy.

There may be more than one way to translate our discussions of such two-sided tests in HEP into the language of severity.  One approach, which Sam Fletcher first explained to me~\citep{fletcher}, is to use absolute values of differences from $\mu=1$ as the ordering ``distance", which leads to summing contributions from two tails.  

Another approach, discussed in SIST (p.\ 154), refers to Cox's taxonomy of tests to decompose the two-sided test into two one-tailed tests, ``the first to examine the possibility that $\mu>\mu_0$, say, the second for $\mu<\mu_0$".  The overall significance level is twice the smaller $p$-value, because of a ``selection effect", and of course, one reports the direction. The severity application here seems to convey the same information that can be communicated in HEP using confidence distributions.

This approach is also discussed in \citet[p.\ 14]{spanos2024}. He defines $H_0$: $\theta = \theta^\ast = 0.5$, and says, ``When there is no reliable information about the potential neighborhood around $\theta^\ast$, one should always use the two-sided framing $H_0$: $\theta = 0.5$ vs.\ $H_1$: $\theta \ne 0.5$."  Then, post-data, having observed $\hat\theta=0.2$, he says that ``the sign and magnitude of [$\hat\theta - \theta^\ast$] indicate that the relevant inferential claim\dots."

\subsubsection{Further measurements of Higgs boson couplings}

\begin{figure}[t]
\begin{center}
\includegraphics[width=0.7\textwidth]{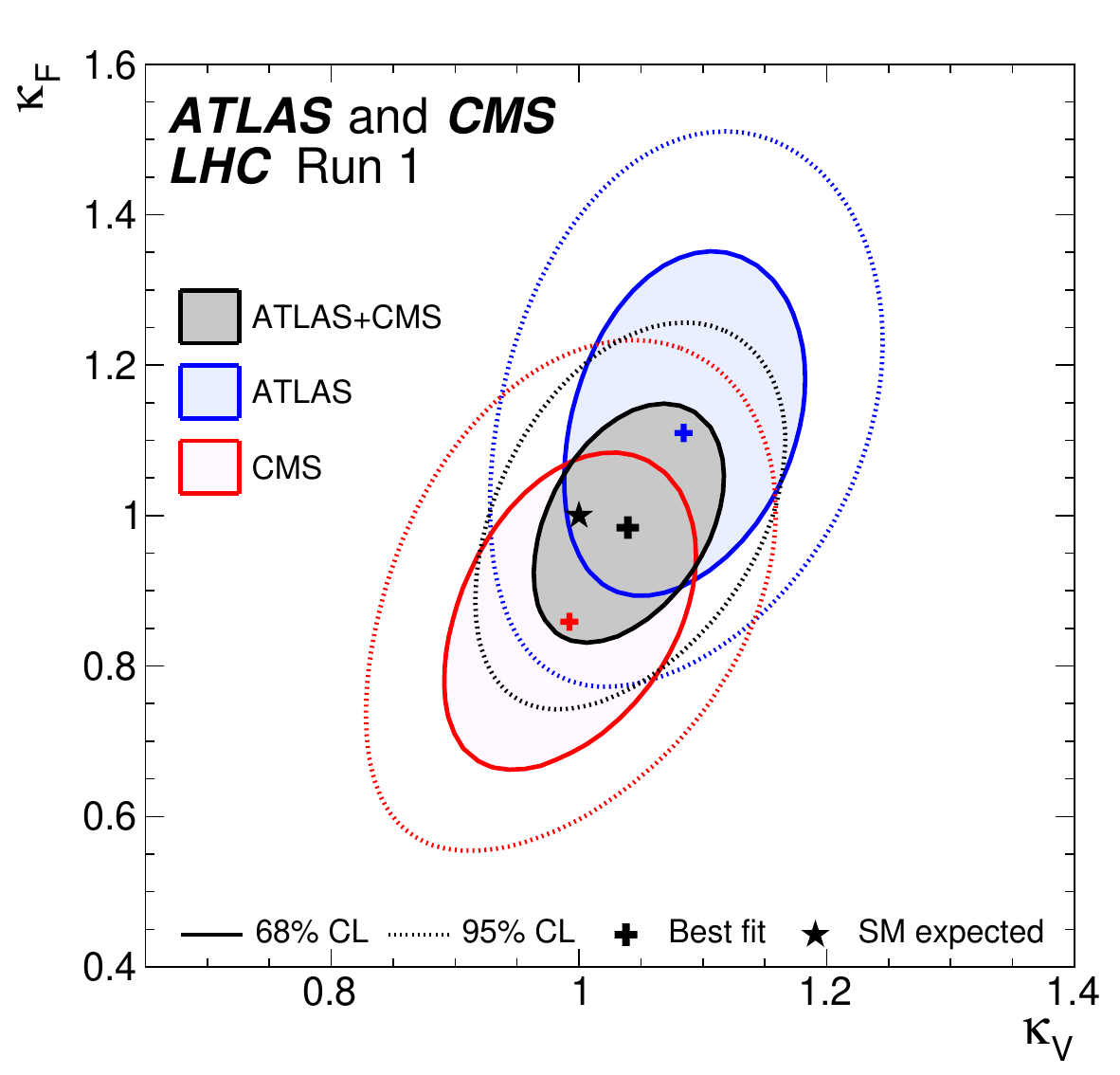}
\end{center}
\caption{Negative log-likelihood contours at~68\% and~95\%~CL in the   ($\kappa_V$,~$\kappa_F$)~plane from the Higgs boson analysis of the combined ATLAS and CMS data sets. Also shown are the contours obtained for each experiment separately. The CL corresponds the claim that the true values are within the respective region. In HEP, we say that values outside the regions are ``excluded'' at the respective CL. There is a long-term program to shrink the regions for the given CL, i.e., to test the claim that the true values lie within smaller regions.  Of course, it would be of great interest to observe regions at very high CL that do not contain the SM values. Figure 26 (upper) of \citet{atlascmscouplings}.}
\label{fig:kVkFpos}
\end{figure}

Figure~\ref{fig:kVkFpos} displays the results of the same suite of analyses (including Higgs boson production measurements) in a complementary way, motivated by the fact that the SM physics equations governing the coupling of the force carriers ($\gamma$, Z, W, known as vector bosons) to the Higgs boson are very different from the physics equations governing the couplings of leptons and quarks (collectively known as fermions) to the Higgs boson.  One inserts an ad-hoc scale factor $\kappa_V$ in the couplings to vector bosons in the equations, and a factor $\kappa_F$ in the equations for fermion couplings, and measures those factors. Figure~\ref{fig:kVkFpos} displays 2D confidence {\em regions} in the plane of $\kappa_F$ vs.\ $\kappa_V$, for two different \cl.  Again, these are two of the confidence regions in a confidence distribution of regions at all \cl.  And again, the results are compatible with the null hypothesis of the SM prediction of $\kappa_F=\kappa_V=1$. 

This was a fantastic beginning to the campaign of measuring couplings, but also motivates the effort (and expense) of obtaining far better precision in the years (and decades) to come.
As wonderful as these early results were, it turns out that many speculative new physics models would alter $\mu=1$ by only a few per cent or less (especially after being constrained by existing LHC results).  Thus, these data cannot distinguish the SM from a lot of the current speculation.  The confidence intervals have since shrunk at the LHC, and will continue to shrink at the upgraded facilities called the High-Luminosity LHC, currently being installed. Meanwhile, there is also a big push for new accelerator(s) to shrink them to be yet smaller, in hopes of discovering new phenomena.

\newcommand{\BR}{\ensuremath{\mathrm{B}}}

\begin{figure}[ht!]
\centering
\vspace{-4mm}
\includegraphics[width=0.7\textwidth]{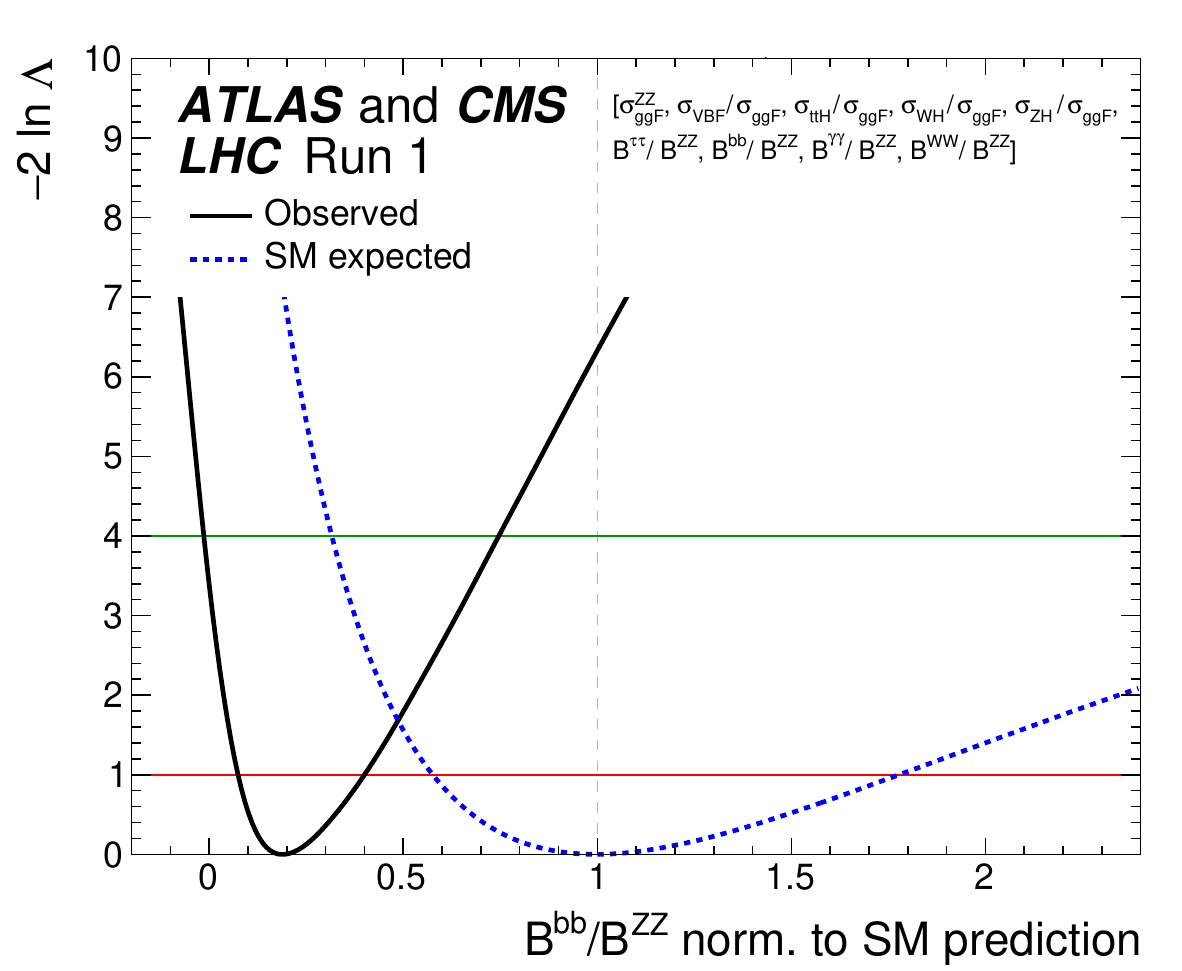}
\caption{Observed (solid line) negative log-likelihood scan of the   $\BR^{bb}/\BR^{ZZ}$ parameter normalized to the corresponding SM~prediction. All the other parameters are also varied in the minimization procedure (profiling). The dashed blue line is the result expected for data conforming exactly to the SM prediction.  The red (green) horizontal line at the $-2\Delta\ln\Lambda$ value of 1 (4) indicates the value of the profile likelihood ratio corresponding to a $1\sigma$ ($2\sigma$) CL~interval for the parameter of interest, assuming the asymptotic $\chi^2$~distribution of the test statistic. The vertical dashed line through unity indicates the SM prediction.  Figure 9 of \citet{atlascmscouplings}.}
\label{fig:rates_bb_likelihoodscan}
\end{figure}

With so many quantities being measured and various ratios of quantities being measured as well (with the advantage that certain systematic effects can cancel in a ratio), it was expected that there might be some outliers, if only because of random fluctuations in the data.  These were scrutinized and reported with additional information. In an interesting case, $\mu$ is the ratio of Higgs boson decays {\em to two $b$ quarks} and Higgs boson decays {\em to two $Z$ bosons}, denoted as the ratio $\BR^{bb}/\BR^{ZZ}$, as usual normalized so that $\mu=1$ is the SM prediction.  The measured $\mu$ departed somewhat from $\mu=1$. A scan of the profile likelihood ratio $\Lambda$ as a function of $\mu$ was computed and displayed as ${-}2\Delta\ln\Lambda$, where $\Delta\ln\Lambda$ is the difference $\ln\Lambda(\mu) - \ln\Lambda(\hat\mu)$.  As seen in Fig.~\ref{fig:rates_bb_likelihoodscan}, the likelihood ratio scan is asymmetric, with the best-fit value being approximately~2.5$\sigma$ below the SM~prediction.

One can read off the confidence distribution for $\mu$ (\pcl for any CL) 
from Fig.~\ref{fig:rates_bb_likelihoodscan} as follows. Since Wilks's Theorem is reasonably valid for these data, and as there is just one degree of freedom ($\mu$) remaining after removing all other parameters via profiling, the quantity plotted has a $\chi^2$ distribution with one degree of freedom.  Thus, reading off the curve at any $\mu$ and taking the square root gives the ``number of $\sigma$'' in an equivalent equal-tailed test of a normal distribution, and thus the p-CL of the dual interval.  The horizontal lines intercept the curve at the endpoints of the 1$\sigma$ (68\%) and 2$\sigma$ (95\%) confidence intervals.  The confidence interval with endpoint on the SM prediction corresponds to about $\sqrt{6.3}\sigma = 2.5\sigma$, i.e., $\pcl\approx99$\%.  One can similarly calculate \pcl for any desired $\mu_1$.
Given the ``multiple trials'' of all the measurements considered (not yet included in the analysis of \pcl), as well as the high threshold in HEP for claiming ``new physics'', this result was put in the category of something to keep an eye on as more data were accumulated. CMS has reported updated measurements in 2026 that still show tension with the SM~\citep{CMScombined2026}.

\section{Summary and final thoughts}
In HEP, regardless of whether or not the null hypothesis is formally rejected according to some rigid application of N-P testing, there is additional data-dependent information to be communicated regarding the compatibility of any chosen value of $\mu$ (or range of $\mu$) with the observed data. Confidence intervals with CL fixed in advance can convey some of this information, but confidence intervals with multiple values of CL (in the extreme, the whole confidence distribution) can convey more.  The confidence interval having critical confidence level \pcl that just barely contains some $\mu_1$ of interest is a quantity of interest that can often be inferred from typical plots in HEP, or computed as scientific interest merits. I see analogies, and in some cases numerical identity, with the concept of severity.

I think that the practice in HEP illustrated in the examples defends against the same ills or fallacies that severity is designed to protect against. I would have to understand severity better before specifying any numerical comparisons, and I hope that others can help with the translation and precise comparison. The underlying {\em philosophy} of what we are doing is talked about much less in HEP and rarely articulated in depth. In my opinion, the tensions between different philosophies are unresolved in HEP, just as they are in the statistics profession, even if we choose the frequentist approach for conveying most experimental results. Tension with Bayesian approaches is particularly evident since our prototype test has a point null hypothesis and a continuous alternative, the situation of the Jeffreys-Lindley paradox.  Thus, I have intentionally kept this paper largely descriptive, rather than prescriptive, regarding practice in HEP.  

\section*{Acknowledgments}
I thank Andrew Brilliant, Kyle Cormier, Andre David, Samuel Fletcher, Louis Lyons, Deborah Mayo,  and members of the CMS Statistics Committee for useful discussions and comments on this and/or earlier drafts. I used ChatGPT to check grammar (suggestions considered and implemented by hand) and to modify the bibtex source for APA style (checked and altered by hand). This work was partially supported by the U.S.\ Department of Energy under Award Number {DE}--{SC}0009937.

\bibliography{cousins_SISTa.bib}{}

\begin{thebibliography}{}
\renewcommand{\doi}[1]{\url{https://doi.org/#1}}
\bibcommenthead

\bibitem [\protect \citeauthoryear {%
Abratenko%
\ \protect \BOthers {.}}{%
Abratenko%
\ \protect \BOthers {.}}{%
{\protect \APACyear {2025}}%
}]{%
microboone2025}
\APACinsertmetastar {%
microboone2025}%
\begin{APACrefauthors}%
Abratenko, P.%
\BCBT {}\ \BOthersPeriod {.}
\end{APACrefauthors}%
\unskip\
\newblock
\APACrefYearMonthDay{2025}{}{}.
\newblock
{\BBOQ}\APACrefatitle {Search for light sterile neutrinos with two neutrino beams at {MicroBooNE}} {Search for light sterile neutrinos with two neutrino beams at {MicroBooNE}}.{\BBCQ}
\newblock
\APACjournalVolNumPages{Nature}{648}{}{64--69,}
\newblock
\begin{APACrefDOI} \doi{10.1038/s41586-025-09757-7} \end{APACrefDOI}
\newblock
{\href{https://arxiv.org/abs/2512.07159}{{arXiv:2512.07159}}}
\newblock
 {[hep-ex]}
\PrintBackRefs{\CurrentBib}

\bibitem [\protect \citeauthoryear {%
Affolder%
\ \protect \BOthers {.}}{%
Affolder%
\ \protect \BOthers {.}}{%
{\protect \APACyear {2000}}%
}]{%
cdffc}
\APACinsertmetastar {%
cdffc}%
\begin{APACrefauthors}%
Affolder, T.%
\BCBT {}\ \BOthersPeriod {.}
\end{APACrefauthors}%
\unskip\
\newblock
\APACrefYearMonthDay{2000}{}{}.
\newblock
{\BBOQ}\APACrefatitle {A measurement of {$\sin(2\beta)$} from {$B \to J/\psi K^0_S$} with the {CDF} detector} {A measurement of {$\sin(2\beta)$} from {$B \to J/\psi K^0_S$} with the {CDF} detector}.{\BBCQ}
\newblock
\APACjournalVolNumPages{Physical Review D}{61}{}{072005,}
\newblock
\begin{APACrefDOI} \doi{10.1103/PhysRevD.61.072005} \end{APACrefDOI}
\newblock
{\href{https://arxiv.org/abs/hep-ex/9909003}{{arXiv:hep-ex/9909003}}}
\newblock
 {[hep-ex]}
\PrintBackRefs{\CurrentBib}

\bibitem [\protect \citeauthoryear {%
{ATLAS and CMS Collaborations}%
, Aad%
\BCBL {}\ \protect \BOthers {.}}{%
{ATLAS and CMS Collaborations}%
\ \protect \BOthers {.}}{%
{\protect \APACyear {2016}}%
}]{%
atlascmscouplings}
\APACinsertmetastar {%
atlascmscouplings}%
\begin{APACrefauthors}%
{ATLAS and CMS Collaborations}%
, Aad, G.%
\BCBL {}\ \BOthersPeriod {.}\end{APACrefauthors}%
\unskip\
\newblock
\APACrefYearMonthDay{2016}{}{}.
\newblock
{\BBOQ}\APACrefatitle {Measurements of the {Higgs} boson production and decay rates and constraints on its couplings from a combined {ATLAS} and {CMS} analysis of the {LHC} pp collision data at $ \sqrt{s}=7 $ and 8 {TeV}} {Measurements of the {Higgs} boson production and decay rates and constraints on its couplings from a combined {ATLAS} and {CMS} analysis of the {LHC} pp collision data at $ \sqrt{s}=7 $ and 8 {TeV}}.{\BBCQ}
\newblock
\APACjournalVolNumPages{JHEP}{08}{}{045,}
\newblock
\begin{APACrefDOI} \doi{10.1007/JHEP08(2016)045} \end{APACrefDOI}
\newblock
{\href{https://arxiv.org/abs/1606.02266}{{arXiv:1606.02266}}}
\newblock
 {[hep-ex]}
\PrintBackRefs{\CurrentBib}

\bibitem [\protect \citeauthoryear {%
{ATLAS and CMS Collaborations, and LHC Higgs Combination Group}%
}{%
{ATLAS and CMS Collaborations, and LHC Higgs Combination Group}%
}{%
{\protect \APACyear {2011}}%
}]{%
CMS-NOTE-2011-005}
\APACinsertmetastar {%
CMS-NOTE-2011-005}%
\begin{APACrefauthors}%
{ATLAS and CMS Collaborations, and LHC Higgs Combination Group}%
\end{APACrefauthors}%
\unskip\
\newblock
\APACrefYearMonthDay{2011}{}{}.
\newblock
\APACrefbtitle {{\textnormal{Procedure for the {LHC} {Higgs} boson search combination in Summer 2011}}} {{\textnormal{Procedure for the {LHC} {Higgs} boson search combination in Summer 2011}}}\ \APACbVolEdTR{}{\BTR{}\ \BNUMS\ CMS-NOTE-2011-005, ATL-PHYS-PUB-2011-11}.
\newblock
\APACrefnote{\url{https://cds.cern.ch/record/1379837}}
\PrintBackRefs{\CurrentBib}

\bibitem [\protect \citeauthoryear {%
{ATLAS Collaboration}%
, Aad%
\BCBL {}\ \protect \BOthers {.}}{%
{ATLAS Collaboration}%
\ \protect \BOthers {.}}{%
{\protect \APACyear {2012}}%
}]{%
atlashiggs}
\APACinsertmetastar {%
atlashiggs}%
\begin{APACrefauthors}%
{ATLAS Collaboration}%
, Aad, G.%
\BCBL {}\ \BOthersPeriod {.}\end{APACrefauthors}%
\unskip\
\newblock
\APACrefYearMonthDay{2012}{}{}.
\newblock
{\BBOQ}\APACrefatitle {Observation of a new particle in the search for the {S}tandard {M}odel {H}iggs boson with the {ATLAS} detector at the {LHC}} {Observation of a new particle in the search for the {S}tandard {M}odel {H}iggs boson with the {ATLAS} detector at the {LHC}}.{\BBCQ}
\newblock
\APACjournalVolNumPages{Physics Letters B}{716}{}{1--29,}
\newblock
\begin{APACrefDOI} \doi{10.1016/j.physletb.2012.08.020} \end{APACrefDOI}
\newblock

\newblock

\PrintBackRefs{\CurrentBib}

\bibitem [\protect \citeauthoryear {%
{CMS Collaboration}%
, Chatrchyan%
\BCBL {}\ \protect \BOthers {.}}{%
{CMS Collaboration}%
\ \protect \BOthers {.}}{%
{\protect \APACyear {2012}}%
{\protect \APACexlab {{\protect \BCnt {1}}}}}]{%
cms_limits_2012}
\APACinsertmetastar {%
cms_limits_2012}%
\begin{APACrefauthors}%
{CMS Collaboration}%
, Chatrchyan, S.%
\BCBL {}\ \BOthersPeriod {.}\end{APACrefauthors}%
\unskip\
\newblock
\APACrefYearMonthDay{2012{\protect \BCnt {1}}}{}{}.
\newblock
{\BBOQ}\APACrefatitle {Combined results of searches for the standard model {H}iggs boson in pp collisions at $\sqrt{s} =7$ {TeV}} {Combined results of searches for the standard model {H}iggs boson in pp collisions at $\sqrt{s} =7$ {TeV}}.{\BBCQ}
\newblock
\APACjournalVolNumPages{Physics Letters B}{710}{}{26--48,}
\newblock
\begin{APACrefDOI} \doi{10.1016/j.physletb.2012.02.064} \end{APACrefDOI}
\newblock
{\href{https://arxiv.org/abs/1202.1488}{{arXiv:1202.1488}}}
\newblock
 {[hep-ex]}
\PrintBackRefs{\CurrentBib}

\bibitem [\protect \citeauthoryear {%
{CMS Collaboration}%
, Chatrchyan%
\BCBL {}\ \protect \BOthers {.}}{%
{CMS Collaboration}%
\ \protect \BOthers {.}}{%
{\protect \APACyear {2012}}%
{\protect \APACexlab {{\protect \BCnt {2}}}}}]{%
cmshiggs}
\APACinsertmetastar {%
cmshiggs}%
\begin{APACrefauthors}%
{CMS Collaboration}%
, Chatrchyan, S.%
\BCBL {}\ \BOthersPeriod {.}\end{APACrefauthors}%
\unskip\
\newblock
\APACrefYearMonthDay{2012{\protect \BCnt {2}}}{}{}.
\newblock
{\BBOQ}\APACrefatitle {Observation of a new boson at a mass of 125 {GeV} with the {CMS} experiment at the {LHC}} {Observation of a new boson at a mass of 125 {GeV} with the {CMS} experiment at the {LHC}}.{\BBCQ}
\newblock
\APACjournalVolNumPages{Physics Letters B}{716}{}{30--61,}
\newblock
\begin{APACrefDOI} \doi{10.1016/j.physletb.2012.08.021} \end{APACrefDOI}
\newblock

\newblock

\PrintBackRefs{\CurrentBib}

\bibitem [\protect \citeauthoryear {%
{CMS Collaboration}%
, Chatrchyan%
\BCBL {}\ \protect \BOthers {.}}{%
{CMS Collaboration}%
\ \protect \BOthers {.}}{%
{\protect \APACyear {2013}}%
}]{%
cmshiggsprop2012}
\APACinsertmetastar {%
cmshiggsprop2012}%
\begin{APACrefauthors}%
{CMS Collaboration}%
, Chatrchyan, S.%
\BCBL {}\ \BOthersPeriod {.}\end{APACrefauthors}%
\unskip\
\newblock
\APACrefYearMonthDay{2013}{}{}.
\newblock
{\BBOQ}\APACrefatitle {Study of the Mass and Spin-Parity of the {H}iggs Boson Candidate via Its Decays to {$Z$} Boson Pairs} {Study of the mass and spin-parity of the {H}iggs boson candidate via its decays to {$Z$} boson pairs}.{\BBCQ}
\newblock
\APACjournalVolNumPages{Physical Review Letters}{110}{}{081803,}
\newblock
\begin{APACrefDOI} \doi{10.1103/PhysRevLett.110.081803} \end{APACrefDOI}
\newblock
\APACrefnote{{Erratum: Physical Review Letters 110, 189901 (2013), \url{https://doi.org/10.1103/PhysRevLett.110.189901}}}
\newblock
{\href{https://arxiv.org/abs/1212.6639}{{arXiv:1212.6639}}}
\newblock
 {[hep-ex]}
\PrintBackRefs{\CurrentBib}

\bibitem [\protect \citeauthoryear {%
{CMS Collaboration}%
, Gevorgyan%
\BCBL {}\ \protect \BOthers {.}}{%
{CMS Collaboration}%
, Gevorgyan%
\BCBL {}\ \protect \BOthers {.}}{%
{\protect \APACyear {2026}}%
}]{%
cpmixtureHIG-25-012}
\APACinsertmetastar {%
cpmixtureHIG-25-012}%
\begin{APACrefauthors}%
{CMS Collaboration}%
, Gevorgyan, A.%
\BCBL {}\ \BOthersPeriod {.}\end{APACrefauthors}%
\unskip\
\newblock
\APACrefYearMonthDay{2026}{}{}.
\newblock
{\BBOQ}\APACrefatitle {Analysis of the {$CP$} structure of the {Y}ukawa coupling between the {H}iggs boson and tau leptons in proton-proton collisions at $\sqrt{s}$ = 13.6 {TeV}} {Analysis of the {$CP$} structure of the {Y}ukawa coupling between the {H}iggs boson and tau leptons in proton-proton collisions at $\sqrt{s}$ = 13.6 {TeV}}.{\BBCQ}
\newblock
\APACjournalVolNumPages{JHEP}{}{}{,}
\newblock
\APACrefnote{\url{https://arxiv.org/abs/2606.03510} \fix{accepted; to be updated}}
\newblock
{\href{https://arxiv.org/abs/2606.03510}{{arXiv:2606.03510}}}
\newblock
 {[hep-ex]}
\PrintBackRefs{\CurrentBib}

\bibitem [\protect \citeauthoryear {%
{CMS Collaboration}%
, Hayrapetyan%
\BCBL {}\ \protect \BOthers {.}}{%
{CMS Collaboration}%
\ \protect \BOthers {.}}{%
{\protect \APACyear {2024}}%
}]{%
combine}
\APACinsertmetastar {%
combine}%
\begin{APACrefauthors}%
{CMS Collaboration}%
, Hayrapetyan, A.%
\BCBL {}\ \BOthersPeriod {.}\end{APACrefauthors}%
\unskip\
\newblock
\APACrefYearMonthDay{2024}{}{}.
\newblock
{\BBOQ}\APACrefatitle {{The CMS statistical analysis and combination tool: Combine}} {{The CMS statistical analysis and combination tool: Combine}}.{\BBCQ}
\newblock
\APACjournalVolNumPages{Computing and Software for Big Science}{8}{}{19,}
\newblock
\begin{APACrefDOI} \doi{10.1007/s41781-024-00121-4} \end{APACrefDOI}
\newblock
{\href{https://arxiv.org/abs/2404.06614}{{arXiv:2404.06614}}}
\newblock
 {[physics.data-an]}
\PrintBackRefs{\CurrentBib}

\bibitem [\protect \citeauthoryear {%
{CMS Collaboration}%
, Hayrapetyan%
\BCBL {}\ \protect \BOthers {.}}{%
{CMS Collaboration}%
, Hayrapetyan%
\BCBL {}\ \protect \BOthers {.}}{%
{\protect \APACyear {2026}}%
}]{%
CMScombined2026}
\APACinsertmetastar {%
CMScombined2026}%
\begin{APACrefauthors}%
{CMS Collaboration}%
, Hayrapetyan, A.%
\BCBL {}\ \BOthersPeriod {.}\end{APACrefauthors}%
\unskip\
\newblock
\APACrefYearMonthDay{2026}{}{}.
\newblock
{\BBOQ}\APACrefatitle {Combined measurements and interpretations of {H}iggs boson production and decay in proton-proton collisions at $\sqrt{s} = 13$ {TeV}} {Combined measurements and interpretations of {H}iggs boson production and decay in proton-proton collisions at $\sqrt{s} = 13$ {TeV}}.{\BBCQ}
\newblock
\APACjournalVolNumPages{Reports on Progress in Physics}{89}{}{087801,}
\newblock
\begin{APACrefDOI} \doi{10.1088/1361-6633/ae72c4} \end{APACrefDOI}
\newblock
{\href{https://arxiv.org/abs/2602.18611}{{arXiv:2602.18611}}}
\newblock
 {[hep-ex]}
\PrintBackRefs{\CurrentBib}

\bibitem [\protect \citeauthoryear {%
R.~Cousins%
}{%
R.~Cousins%
}{%
{\protect \APACyear {2018}}%
}]{%
cousins-slides}
\APACinsertmetastar {%
cousins-slides}%
\begin{APACrefauthors}%
Cousins, R.%
\end{APACrefauthors}%
\unskip\
\newblock
\APACrefYearMonthDay{2018}{}{}.
\newblock
\APACrefbtitle {{\textnormal{Lectures on Statistics in Theory: Prelude to Statistics in Practice}}.} {{\textnormal{Lectures on Statistics in Theory: Prelude to Statistics in Practice}}.}
\newblock
\APACrefnote{Writeup of lectures at 2018 Hadron Collider Physics Summer School at Fermilab. Slides at \url{https://www.physics.ucla.edu/~cousins/stats/theoryslides}~. \url{https://arxiv.org/abs/1807.05996}}
\PrintBackRefs{\CurrentBib}

\bibitem [\protect \citeauthoryear {%
R.~Cousins%
, Mumford%
, Tucker%
\BCBL {}\ \BBA {} Valuev%
}{%
R.~Cousins%
\ \protect \BOthers {.}}{%
{\protect \APACyear {2005}}%
}]{%
cousinsspin}
\APACinsertmetastar {%
cousinsspin}%
\begin{APACrefauthors}%
Cousins, R.%
, Mumford, J.%
, Tucker, J.%
\BCBL {} Valuev, V.%
\end{APACrefauthors}%
\unskip\
\newblock
\APACrefYearMonthDay{2005}{}{}.
\newblock
{\BBOQ}\APACrefatitle {Spin discrimination of new heavy resonances at the {LHC}} {Spin discrimination of new heavy resonances at the {LHC}}.{\BBCQ}
\newblock
\APACjournalVolNumPages{JHEP}{11}{}{046,}
\newblock
\begin{APACrefDOI} \doi{10.1088/1126-6708/2005/11/046} \end{APACrefDOI}
\newblock

\newblock

\PrintBackRefs{\CurrentBib}

\bibitem [\protect \citeauthoryear {%
R.D.~Cousins%
}{%
R.D.~Cousins%
}{%
{\protect \APACyear {2011}}%
{\protect \APACexlab {{\protect \BCnt {1}}}}}]{%
cousinsvirtual}
\APACinsertmetastar {%
cousinsvirtual}%
\begin{APACrefauthors}%
Cousins, R.D.%
\end{APACrefauthors}%
\unskip\
\newblock
\APACrefYearMonthDay{2011{\protect \BCnt {1}}}{}{}.
\newblock
\APACrefbtitle {{\textnormal{Bayes, {F}isher, {N}eyman, Neutrino Masses, and the {LHC}}}.} {{\textnormal{Bayes, {F}isher, {N}eyman, Neutrino Masses, and the {LHC}}}.}
\newblock
\APACrefnote{\url{https://www.physics.ucla.edu/~cousins/stats/cousins_bounded_gaussian_virtual_talk_12sep2011.pdf}~}
\PrintBackRefs{\CurrentBib}

\bibitem [\protect \citeauthoryear {%
R.D.~Cousins%
}{%
R.D.~Cousins%
}{%
{\protect \APACyear {2011}}%
{\protect \APACexlab {{\protect \BCnt {2}}}}}]{%
cousins2011}
\APACinsertmetastar {%
cousins2011}%
\begin{APACrefauthors}%
Cousins, R.D.%
\end{APACrefauthors}%
\unskip\
\newblock
\APACrefYearMonthDay{2011{\protect \BCnt {2}}}{}{}.
\newblock
\APACrefbtitle {{\textnormal{Negatively biased relevant subsets induced by the most-powerful one-sided upper confidence limits for a bounded physical parameter}}.} {{\textnormal{Negatively biased relevant subsets induced by the most-powerful one-sided upper confidence limits for a bounded physical parameter}}.}
\newblock
\APACrefnote{\url{https://arxiv.org/abs/1109.2023}}
\PrintBackRefs{\CurrentBib}

\bibitem [\protect \citeauthoryear {%
R.D.~Cousins%
}{%
R.D.~Cousins%
}{%
{\protect \APACyear {2017}}%
}]{%
cousinsJL}
\APACinsertmetastar {%
cousinsJL}%
\begin{APACrefauthors}%
Cousins, R.D.%
\end{APACrefauthors}%
\unskip\
\newblock
\APACrefYearMonthDay{2017}{}{}.
\newblock
{\BBOQ}\APACrefatitle {The {J}effreys--{L}indley paradox and discovery criteria in high energy physics} {The {J}effreys--{L}indley paradox and discovery criteria in high energy physics}.{\BBCQ}
\newblock
\APACjournalVolNumPages{Synthese}{194}{1}{395--432,}
\newblock
\begin{APACrefDOI} \doi{10.1007/s11229-014-0525-z} \end{APACrefDOI}
\newblock
\APACrefnote{{E}rratum \url{https://doi.org/10.1007/s11229-015-0687-3}}
\newblock
{\href{https://arxiv.org/abs/1310.3791}{{arXiv:1310.3791}}}
\newblock
 {[physics.data-an]}
\PrintBackRefs{\CurrentBib}

\bibitem [\protect \citeauthoryear {%
R.D.~Cousins%
\ \BBA {} Wasserman%
}{%
R.D.~Cousins%
\ \BBA {} Wasserman%
}{%
{\protect \APACyear {2024}}%
}]{%
cousinswasserman}
\APACinsertmetastar {%
cousinswasserman}%
\begin{APACrefauthors}%
Cousins, R.D.%
\BCBT {}\ \BBA {} Wasserman, L.%
\end{APACrefauthors}%
\unskip\
\newblock
\APACrefYearMonthDay{2024}{}{}.
\newblock
\APACrefbtitle {{\textnormal{{PHYSTAT} informal review: {M}arginalizing versus profiling of nuisance parameters}}.} {{\textnormal{{PHYSTAT} informal review: {M}arginalizing versus profiling of nuisance parameters}}.}
\newblock
\APACrefnote{\url{https://arxiv.org/abs/2404.17180}}
\PrintBackRefs{\CurrentBib}

\bibitem [\protect \citeauthoryear {%
Cowan%
, Cranmer%
, Gross%
\BCBL {}\ \BBA {} Vitells%
}{%
Cowan%
\ \protect \BOthers {.}}{%
{\protect \APACyear {2011}}%
}]{%
pcl}
\APACinsertmetastar {%
pcl}%
\begin{APACrefauthors}%
Cowan, G.%
, Cranmer, K.%
, Gross, E.%
\BCBL {} Vitells, O.%
\end{APACrefauthors}%
\unskip\
\newblock
\APACrefYearMonthDay{2011}{}{}.
\newblock
\APACrefbtitle {Power-Constrained Limits.} {Power-constrained limits.}
\newblock
\APACrefnote{\url{https://arxiv.org/abs/1105.3166}}
\PrintBackRefs{\CurrentBib}

\bibitem [\protect \citeauthoryear {%
Cox%
}{%
Cox%
}{%
{\protect \APACyear {1958}}%
}]{%
cox1958}
\APACinsertmetastar {%
cox1958}%
\begin{APACrefauthors}%
Cox, D.R.%
\end{APACrefauthors}%
\unskip\
\newblock
\APACrefYearMonthDay{1958}{}{}.
\newblock
{\BBOQ}\APACrefatitle {Some Problems Connected with Statistical Inference} {Some problems connected with statistical inference}.{\BBCQ}
\newblock
\APACjournalVolNumPages{The Annals of Mathematical Statistics}{29}{2}{357--372,}
\newblock
\begin{APACrefDOI} \doi{10.1214/aoms/1177706618} \end{APACrefDOI}
\newblock

\newblock

\PrintBackRefs{\CurrentBib}

\bibitem [\protect \citeauthoryear {%
Demortier%
\ \BBA {} Lyons%
}{%
Demortier%
\ \BBA {} Lyons%
}{%
{\protect \APACyear {2014}}%
}]{%
demortierlyons}
\APACinsertmetastar {%
demortierlyons}%
\begin{APACrefauthors}%
Demortier, L.%
\BCBT {}\ \BBA {} Lyons, L.%
\end{APACrefauthors}%
\unskip\
\newblock
\APACrefYearMonthDay{2014}{}{}.
\newblock
\APACrefbtitle {{\textnormal{Testing hypotheses in particle physics: {P}lots of $p_{0}$ Versus $p_{1}$}}.} {{\textnormal{Testing hypotheses in particle physics: {P}lots of $p_{0}$ Versus $p_{1}$}}.}
\newblock
\APACrefnote{\url{https://arxiv.org/abs/1408.6123}}
\PrintBackRefs{\CurrentBib}

\bibitem [\protect \citeauthoryear {%
Eadie%
, Drijard%
, James%
, Roos%
\BCBL {}\ \BBA {} Sadoulet%
}{%
Eadie%
\ \protect \BOthers {.}}{%
{\protect \APACyear {1971}}%
}]{%
eadie1971}
\APACinsertmetastar {%
eadie1971}%
\begin{APACrefauthors}%
Eadie, W.T.%
, Drijard, D.%
, James, F.E.%
, Roos, M.%
\BCBL {} Sadoulet, B.%
\end{APACrefauthors}%
\unskip\
\newblock
\APACrefYear{1971}.
\newblock
\APACrefbtitle {Statistical Methods in Experimental Physics} {Statistical methods in experimental physics}\ (\PrintOrdinal{1st}\ \BEd).
\newblock
\APACaddressPublisher{}{North Holland}.
\PrintBackRefs{\CurrentBib}

\bibitem [\protect \citeauthoryear {%
Efron%
\ \BBA {} Tibshirani%
}{%
Efron%
\ \BBA {} Tibshirani%
}{%
{\protect \APACyear {1998}}%
}]{%
efron1998}
\APACinsertmetastar {%
efron1998}%
\begin{APACrefauthors}%
Efron, B.%
\BCBT {}\ \BBA {} Tibshirani, R.%
\end{APACrefauthors}%
\unskip\
\newblock
\APACrefYearMonthDay{1998}{}{}.
\newblock
{\BBOQ}\APACrefatitle {The Problem of Regions} {The problem of regions}.{\BBCQ}
\newblock
\APACjournalVolNumPages{The Annals of Statistics}{26}{5}{1687--1718,}
\newblock
\begin{APACrefDOI} \doi{10.1214/aos/1024691353} \end{APACrefDOI}
\newblock

\newblock

\PrintBackRefs{\CurrentBib}

\bibitem [\protect \citeauthoryear {%
Feldman%
\ \BBA {} Cousins%
}{%
Feldman%
\ \BBA {} Cousins%
}{%
{\protect \APACyear {1998}}%
}]{%
feldman1998}
\APACinsertmetastar {%
feldman1998}%
\begin{APACrefauthors}%
Feldman, G.J.%
\BCBT {}\ \BBA {} Cousins, R.D.%
\end{APACrefauthors}%
\unskip\
\newblock
\APACrefYearMonthDay{1998}{}{}.
\newblock
{\BBOQ}\APACrefatitle {Unified Approach to the Classical Statistical Analysis of Small Signals} {Unified approach to the classical statistical analysis of small signals}.{\BBCQ}
\newblock
\APACjournalVolNumPages{Physical Review D}{57}{}{3873--3889,}
\newblock
\begin{APACrefDOI} \doi{10.1103/PhysRevD.57.3873} \end{APACrefDOI}
\newblock
{\href{https://arxiv.org/abs/physics/9711021}{{arXiv:physics/9711021}}}
\newblock

\PrintBackRefs{\CurrentBib}

\bibitem [\protect \citeauthoryear {%
Fletcher%
}{%
Fletcher%
}{%
{\protect \APACyear {2020}}%
}]{%
fletcher}
\APACinsertmetastar {%
fletcher}%
\begin{APACrefauthors}%
Fletcher, S.C.%
\end{APACrefauthors}%
\unskip\
\newblock
\APACrefYearMonthDay{2020}{}{}.
\newblock
\APACrefbtitle {{\textnormal{Severity and its potential applications in particle physics data analysis}}.} {{\textnormal{Severity and its potential applications in particle physics data analysis}}.}
\newblock
\APACrefnote{Talk at Carnegie Mellon Dept.\ of Statistics \& Data Science, March 20, 2020}
\PrintBackRefs{\CurrentBib}

\bibitem [\protect \citeauthoryear {%
Hoenig%
\ \BBA {} Heisey%
}{%
Hoenig%
\ \BBA {} Heisey%
}{%
{\protect \APACyear {2001}}%
}]{%
HoenigHeisey2001}
\APACinsertmetastar {%
HoenigHeisey2001}%
\begin{APACrefauthors}%
Hoenig, J.M.%
\BCBT {}\ \BBA {} Heisey, D.M.%
\end{APACrefauthors}%
\unskip\
\newblock
\APACrefYearMonthDay{2001}{}{}.
\newblock
{\BBOQ}\APACrefatitle {The Abuse of Power: The Pervasive Fallacy of Power Calculations for Data Analysis} {The abuse of power: The pervasive fallacy of power calculations for data analysis}.{\BBCQ}
\newblock
\APACjournalVolNumPages{The American Statistician}{55}{1}{19--24,}
\newblock
\begin{APACrefDOI} \doi{10.1198/000313001300339897} \end{APACrefDOI}
\newblock

\newblock

\PrintBackRefs{\CurrentBib}

\bibitem [\protect \citeauthoryear {%
James%
}{%
James%
}{%
{\protect \APACyear {1980}}%
}]{%
james1980}
\APACinsertmetastar {%
james1980}%
\begin{APACrefauthors}%
James, F.%
\end{APACrefauthors}%
\unskip\
\newblock
\APACrefYearMonthDay{1980}{}{}.
\newblock
{\BBOQ}\APACrefatitle {Interpretation of the Shape of the Likelihood Function around Its Minimum} {Interpretation of the shape of the likelihood function around its minimum}.{\BBCQ}
\newblock
\APACjournalVolNumPages{Computer Physics Communications}{20}{1}{29--35,}
\newblock
\begin{APACrefDOI} \doi{10.1016/0010-4655(80)90103-4} \end{APACrefDOI}
\newblock

\newblock

\PrintBackRefs{\CurrentBib}

\bibitem [\protect \citeauthoryear {%
James%
\ \BBA {} Roos%
}{%
James%
\ \BBA {} Roos%
}{%
{\protect \APACyear {1975}}%
}]{%
JamesMinuit1975}
\APACinsertmetastar {%
JamesMinuit1975}%
\begin{APACrefauthors}%
James, F.%
\BCBT {}\ \BBA {} Roos, M.%
\end{APACrefauthors}%
\unskip\
\newblock
\APACrefYearMonthDay{1975}{}{}.
\newblock
{\BBOQ}\APACrefatitle {Minuit: {A} System for Function Minimization and Analysis of the Parameter Errors and Correlations} {Minuit: {A} system for function minimization and analysis of the parameter errors and correlations}.{\BBCQ}
\newblock
\APACjournalVolNumPages{Computer Physics Communications}{10}{}{343--367,}
\newblock
\begin{APACrefDOI} \doi{10.1016/0010-4655(75)90039-9} \end{APACrefDOI}
\newblock

\newblock

\PrintBackRefs{\CurrentBib}

\bibitem [\protect \citeauthoryear {%
Lehmann%
\ \BBA {} Romano%
}{%
Lehmann%
\ \BBA {} Romano%
}{%
{\protect \APACyear {2005}}%
}]{%
lehmann}
\APACinsertmetastar {%
lehmann}%
\begin{APACrefauthors}%
Lehmann, E.L.%
\BCBT {}\ \BBA {} Romano, J.P.%
\end{APACrefauthors}%
\unskip\
\newblock
\APACrefYear{2005}.
\newblock
\APACrefbtitle {Testing Statistical Hypotheses} {Testing statistical hypotheses}\ (\PrintOrdinal{3rd}\ \BEd).
\newblock
\APACaddressPublisher{}{Springer}.
\PrintBackRefs{\CurrentBib}

\bibitem [\protect \citeauthoryear {%
Mandelkern%
}{%
Mandelkern%
}{%
{\protect \APACyear {2002}}%
}]{%
mandelkern2002}
\APACinsertmetastar {%
mandelkern2002}%
\begin{APACrefauthors}%
Mandelkern, M.%
\end{APACrefauthors}%
\unskip\
\newblock
\APACrefYearMonthDay{2002}{}{}.
\newblock
{\BBOQ}\APACrefatitle {Setting Confidence Intervals for Bounded Parameters} {Setting confidence intervals for bounded parameters}.{\BBCQ}
\newblock
\APACjournalVolNumPages{Statistical Science}{17}{2}{149--172,}
\newblock
\begin{APACrefDOI} \doi{10.1214/ss/1030550859} \end{APACrefDOI}
\newblock
\APACrefnote{With comments by five statisticians and rejoinder}
\newblock

\newblock

\PrintBackRefs{\CurrentBib}

\bibitem [\protect \citeauthoryear {%
Mayo%
}{%
Mayo%
}{%
{\protect \APACyear {2018}}%
}]{%
SIST}
\APACinsertmetastar {%
SIST}%
\begin{APACrefauthors}%
Mayo, D.G.%
\end{APACrefauthors}%
\unskip\
\newblock
\APACrefYear{2018}.
\newblock
\APACrefbtitle {Statistical Inference as Severe Testing: How to Get Beyond the Statistics Wars} {Statistical inference as severe testing: How to get beyond the statistics wars}.
\newblock
\APACaddressPublisher{}{Cambridge University Press}.
\PrintBackRefs{\CurrentBib}

\bibitem [\protect \citeauthoryear {%
Mayo%
\ \BBA {} Spanos%
}{%
Mayo%
\ \BBA {} Spanos%
}{%
{\protect \APACyear {2006}}%
}]{%
mayospanos2006}
\APACinsertmetastar {%
mayospanos2006}%
\begin{APACrefauthors}%
Mayo, D.G.%
\BCBT {}\ \BBA {} Spanos, A.%
\end{APACrefauthors}%
\unskip\
\newblock
\APACrefYearMonthDay{2006}{}{}.
\newblock
{\BBOQ}\APACrefatitle {Severe Testing as a Basic Concept in a {N}eyman-{P}earson Philosophy of Induction} {Severe testing as a basic concept in a {N}eyman-{P}earson philosophy of induction}.{\BBCQ}
\newblock
\APACjournalVolNumPages{The British Journal for the Philosophy of Science}{57}{}{323--357,}
\newblock
\begin{APACrefDOI} \doi{10.1093/bjps/axl003} \end{APACrefDOI}
\newblock

\newblock

\PrintBackRefs{\CurrentBib}

\bibitem [\protect \citeauthoryear {%
Mayo%
\ \BBA {} Spanos%
}{%
Mayo%
\ \BBA {} Spanos%
}{%
{\protect \APACyear {2011}}%
}]{%
mayospanos2011error}
\APACinsertmetastar {%
mayospanos2011error}%
\begin{APACrefauthors}%
Mayo, D.G.%
\BCBT {}\ \BBA {} Spanos, A.%
\end{APACrefauthors}%
\unskip\
\newblock
\APACrefYearMonthDay{2011}{}{}.
\newblock
{\BBOQ}\APACrefatitle {Error Statistics} {Error statistics}.{\BBCQ}
\newblock
 P.S.~Bandyopadhyay\ \BBA {} M.R.~Forster\ (\BEDS), \APACrefbtitle {Handbook of the Philosophy of Science: Philosophy of Statistics} {Handbook of the philosophy of science: Philosophy of statistics}\ (\BVOL~7, \BPGS\ 153--198).
\newblock
\APACaddressPublisher{}{Elsevier B.V.}
\PrintBackRefs{\CurrentBib}

\bibitem [\protect \citeauthoryear {%
Neyman%
}{%
Neyman%
}{%
{\protect \APACyear {1937}}%
}]{%
neyman1937}
\APACinsertmetastar {%
neyman1937}%
\begin{APACrefauthors}%
Neyman, J.%
\end{APACrefauthors}%
\unskip\
\newblock
\APACrefYearMonthDay{1937}{}{}.
\newblock
{\BBOQ}\APACrefatitle {Outline of a Theory of Statistical Estimation Based on the Classical Theory of Probability} {Outline of a theory of statistical estimation based on the classical theory of probability}.{\BBCQ}
\newblock
\APACjournalVolNumPages{Philosophical Transactions of the Royal Society of London. Series A, Mathematical and Physical Sciences}{236}{}{333--380,}
\newblock
\begin{APACrefDOI} \doi{10.1098/rsta.1937.0005} \end{APACrefDOI}
\newblock

\newblock

\PrintBackRefs{\CurrentBib}

\bibitem [\protect \citeauthoryear {%
{Particle Data Group}%
, Takahashi%
\BCBL {}\ \protect \BOthers {.}}{%
{Particle Data Group}%
\ \protect \BOthers {.}}{%
{\protect \APACyear {2026}}%
}]{%
pdg2026}
\APACinsertmetastar {%
pdg2026}%
\begin{APACrefauthors}%
{Particle Data Group}%
, Takahashi, F.%
\BCBL {}\ \BOthersPeriod {.}\end{APACrefauthors}%
\unskip\
\newblock
\APACrefYearMonthDay{2026}{}{}.
\newblock
{\BBOQ}\APACrefatitle {{Review of Particle Physics}} {{Review of Particle Physics}}.{\BBCQ}
\newblock
\APACjournalVolNumPages{International Journal of Modern Physics A}{41}{}{2630011,}
\newblock
\begin{APACrefDOI} \doi{10.1142/S0217751X26300115} \end{APACrefDOI}
\newblock
\APACrefnote{\url{https://pdg.lbl.gov/}~}
\newblock

\newblock

\PrintBackRefs{\CurrentBib}

\bibitem [\protect \citeauthoryear {%
Pearson%
}{%
Pearson%
}{%
{\protect \APACyear {1947}}%
}]{%
pearson1947ordering}
\APACinsertmetastar {%
pearson1947ordering}%
\begin{APACrefauthors}%
Pearson, E.S.%
\end{APACrefauthors}%
\unskip\
\newblock
\APACrefYearMonthDay{1947}{}{}.
\newblock
{\BBOQ}\APACrefatitle {The choice of statistical tests illustrated on the interpretation of data classed in a $2\times2$ table} {The choice of statistical tests illustrated on the interpretation of data classed in a $2\times2$ table}.{\BBCQ}
\newblock
\APACjournalVolNumPages{Biometrika}{34}{1--2}{139--167,}
\newblock
\begin{APACrefDOI} \doi{10.1093/biomet/34.1-2.139} \end{APACrefDOI}
\newblock

\newblock

\PrintBackRefs{\CurrentBib}

\bibitem [\protect \citeauthoryear {%
{PHYSTAT}%
}{%
{PHYSTAT}%
}{%
{\protect \APACyear {{\protect \bibnodate {}}}}%
}]{%
phystat}
\APACinsertmetastar {%
phystat}%
\begin{APACrefauthors}%
{PHYSTAT}%
\end{APACrefauthors}%
\unskip\
\newblock
\APACrefYearMonthDay{{\protect \bibnodate {}}}{}{}.
\newblock
\APACrefbtitle {{PHYSTAT}.} {{PHYSTAT}.}
\newblock
\APACrefnote{\url{https://phystat.github.io/Website/}}
\PrintBackRefs{\CurrentBib}

\bibitem [\protect \citeauthoryear {%
Read%
}{%
Read%
}{%
{\protect \APACyear {2002}}%
}]{%
readjphysg}
\APACinsertmetastar {%
readjphysg}%
\begin{APACrefauthors}%
Read, A.L.%
\end{APACrefauthors}%
\unskip\
\newblock
\APACrefYearMonthDay{2002}{}{}.
\newblock
{\BBOQ}\APACrefatitle {Presentation of search results: The {CL$_{\text{s}}$} technique} {Presentation of search results: The {CL$_{\text{s}}$} technique}.{\BBCQ}
\newblock
\APACjournalVolNumPages{Journal of Physics G: Nuclear and Particle Physics}{28}{}{2693--2704,}
\newblock
\begin{APACrefDOI} \doi{10.1088/0954-3899/28/10/313} \end{APACrefDOI}
\newblock

\newblock

\PrintBackRefs{\CurrentBib}

\bibitem [\protect \citeauthoryear {%
Rolke%
}{%
Rolke%
}{%
{\protect \APACyear {2000}}%
}]{%
rolkeclk}
\APACinsertmetastar {%
rolkeclk}%
\begin{APACrefauthors}%
Rolke, W.%
\end{APACrefauthors}%
\unskip\
\newblock
\APACrefYearMonthDay{2000}{}{}.
\newblock
\APACrefbtitle {Confidence intervals and upper bounds for small signals in the presence of noise.} {Confidence intervals and upper bounds for small signals in the presence of noise.}
\newblock
\APACrefnote{Talk at Workshop on Confidence Limits, Fermilab, 27-28 March 2000. \url{https://conferences.fnal.gov/cl2k/}.}
\PrintBackRefs{\CurrentBib}

\bibitem [\protect \citeauthoryear {%
Segreto%
}{%
Segreto%
}{%
{\protect \APACyear {2013}}%
}]{%
Segreto:2013esa}
\APACinsertmetastar {%
Segreto:2013esa}%
\begin{APACrefauthors}%
Segreto, E.%
\end{APACrefauthors}%
\unskip\
\newblock
\APACrefYearMonthDay{2013}{}{}.
\newblock
{\BBOQ}\APACrefatitle {{Experimental search for the LSND anomaly with the ICARUS LAr-TPC detector in the CNGS beam}} {{Experimental search for the LSND anomaly with the ICARUS LAr-TPC detector in the CNGS beam}}.{\BBCQ}
\newblock
\APACjournalVolNumPages{Journal of Physics: Conference Series}{447}{}{012064,}
\newblock
\begin{APACrefDOI} \doi{10.1088/1742-6596/447/1/012064} \end{APACrefDOI}
\newblock

\newblock

\PrintBackRefs{\CurrentBib}

\bibitem [\protect \citeauthoryear {%
Senn%
}{%
Senn%
}{%
{\protect \APACyear {2021}}%
}]{%
sennbook2021}
\APACinsertmetastar {%
sennbook2021}%
\begin{APACrefauthors}%
Senn, S.%
\end{APACrefauthors}%
\unskip\
\newblock
\APACrefYear{2021}.
\newblock
\APACrefbtitle {Statistical Issues in Drug Development} {Statistical issues in drug development}\ (\PrintOrdinal{3rd}\ \BEd).
\newblock
\APACaddressPublisher{}{Wiley}.
\PrintBackRefs{\CurrentBib}

\bibitem [\protect \citeauthoryear {%
Spanos%
}{%
Spanos%
}{%
{\protect \APACyear {2014}}%
}]{%
spanos2014}
\APACinsertmetastar {%
spanos2014}%
\begin{APACrefauthors}%
Spanos, A.%
\end{APACrefauthors}%
\unskip\
\newblock
\APACrefYearMonthDay{2014}{}{}.
\newblock
{\BBOQ}\APACrefatitle {Recurring controversies about {$P$} values and confidence intervals revisited} {Recurring controversies about {$P$} values and confidence intervals revisited}.{\BBCQ}
\newblock
\APACjournalVolNumPages{Ecology}{95}{3}{645--651,}
\newblock
\begin{APACrefDOI} \doi{10.1890/13-1291.1} \end{APACrefDOI}
\newblock

\newblock

\PrintBackRefs{\CurrentBib}

\bibitem [\protect \citeauthoryear {%
Spanos%
}{%
Spanos%
}{%
{\protect \APACyear {2024}}%
}]{%
spanos2024}
\APACinsertmetastar {%
spanos2024}%
\begin{APACrefauthors}%
Spanos, A.%
\end{APACrefauthors}%
\unskip\
\newblock
\APACrefYearMonthDay{2024}{}{}.
\newblock
{\BBOQ}\APACrefatitle {How the Post-Data Severity Converts Testing Results into Evidence for or against Pertinent Inferential Claims} {How the post-data severity converts testing results into evidence for or against pertinent inferential claims}.{\BBCQ}
\newblock
\APACjournalVolNumPages{Entropy}{26}{}{95,}
\newblock
\begin{APACrefDOI} \doi{10.3390/e26010095} \end{APACrefDOI}
\newblock

\newblock

\PrintBackRefs{\CurrentBib}

\bibitem [\protect \citeauthoryear {%
Staley%
}{%
Staley%
}{%
{\protect \APACyear {2014}}%
}]{%
staley}
\APACinsertmetastar {%
staley}%
\begin{APACrefauthors}%
Staley, K.W.%
\end{APACrefauthors}%
\unskip\
\newblock
\APACrefYear{2014}.
\newblock
\APACrefbtitle {An Introduction to the Philosophy of Science} {An introduction to the philosophy of science}.
\newblock
\APACaddressPublisher{}{Cambridge University Press}.
\PrintBackRefs{\CurrentBib}

\bibitem [\protect \citeauthoryear {%
Stuart%
, Ord%
\BCBL {}\ \BBA {} Arnold%
}{%
Stuart%
\ \protect \BOthers {.}}{%
{\protect \APACyear {1999}}%
}]{%
kendall1999}
\APACinsertmetastar {%
kendall1999}%
\begin{APACrefauthors}%
Stuart, A.%
, Ord, K.%
\BCBL {} Arnold, S.%
\end{APACrefauthors}%
\unskip\
\newblock
\APACrefYear{1999}.
\newblock
\APACrefbtitle {Kendall's Advanced Theory of Statistics: Classical Inference and the Linear Model} {Kendall's advanced theory of statistics: Classical inference and the linear model}\ (\PrintOrdinal{6th}\ \BEd, \BVOL~2A).
\newblock
\APACaddressPublisher{}{Arnold}.
\newblock
\APACrefnote{and earlier editions by Kendall and Stuart.}
\PrintBackRefs{\CurrentBib}

\bibitem [\protect \citeauthoryear {%
Thompson%
}{%
Thompson%
}{%
{\protect \APACyear {2007}}%
}]{%
thompson2007}
\APACinsertmetastar {%
thompson2007}%
\begin{APACrefauthors}%
Thompson, B.%
\end{APACrefauthors}%
\unskip\
\newblock
\APACrefYear{2007}.
\newblock
\APACrefbtitle {The Nature of Statistical Evidence ({L}ecture {N}otes in {S}tatistics, Vol. 189)} {The nature of statistical evidence ({L}ecture {N}otes in {S}tatistics, vol. 189)}.
\newblock
\APACaddressPublisher{}{Springer}.
\newblock
\APACrefnote{\url{https://doi.org/10.1007/978-0-387-40054-9}}
\PrintBackRefs{\CurrentBib}

\bibitem [\protect \citeauthoryear {%
{van Dyk}%
}{%
{van Dyk}%
}{%
{\protect \APACyear {2014}}%
}]{%
vandyk2014}
\APACinsertmetastar {%
vandyk2014}%
\begin{APACrefauthors}%
{van Dyk}, D.A.%
\end{APACrefauthors}%
\unskip\
\newblock
\APACrefYearMonthDay{2014}{}{}.
\newblock
{\BBOQ}\APACrefatitle {The Role of Statistics in the Discovery of a {H}iggs Boson} {The role of statistics in the discovery of a {H}iggs boson}.{\BBCQ}
\newblock
\APACjournalVolNumPages{Annual Review of Statistics and Its Application}{1}{}{41--59,}
\newblock
\begin{APACrefDOI} \doi{10.1146/annurev-statistics-062713-085841} \end{APACrefDOI}
\newblock

\newblock

\PrintBackRefs{\CurrentBib}

\bibitem [\protect \citeauthoryear {%
Wilks%
}{%
Wilks%
}{%
{\protect \APACyear {1938}}%
}]{%
wilks38}
\APACinsertmetastar {%
wilks38}%
\begin{APACrefauthors}%
Wilks, S.S.%
\end{APACrefauthors}%
\unskip\
\newblock
\APACrefYearMonthDay{1938}{}{}.
\newblock
{\BBOQ}\APACrefatitle {The Large-Sample Distribution of the Likelihood Ratio for Testing Composite Hypotheses} {The large-sample distribution of the likelihood ratio for testing composite hypotheses}.{\BBCQ}
\newblock
\APACjournalVolNumPages{The Annals of Mathematical Statistics}{9}{1}{60--62,}
\newblock
\begin{APACrefDOI} \doi{10.1214/aoms/1177732360} \end{APACrefDOI}
\newblock

\newblock

\PrintBackRefs{\CurrentBib}

\end{thebibliography}
 
\end{document}